\documentclass[12pt]{iopart}

\usepackage{iopams}  
\usepackage{graphicx}
\usepackage{epstopdf}
\usepackage{afterpage}
\usepackage{braket}
\usepackage{comment}
\usepackage{ulem}
\usepackage{color}
\newcommand{\average}[1]{\ensuremath{\langle#1\rangle} }

\begin{document}
\title{Heat transport via a local two-state system near thermal equilibrium}
\author{Tsuyoshi Yamamoto$^1$, Masanari Kato$^1$, Takeo Kato$^1$ and Keiji Saito$^2$}
\address{$^1$ Institute for Solid State Physics, the University of Tokyo, Kashiwa, Chiba 277-8581, Japan}
\address{$^2$ Department of Physics, Keio University, Yokohama 223-8522, Japan}
\ead{kato@issp.u-tokyo.ac.jp}
\vspace{10pt}
\begin{indented}
\item[]March 2018
\end{indented}

\begin{abstract}
Heat transport in spin-boson systems near the thermal equilibrium is systematically investigated.
An asymptotically exact expression for the thermal conductance in a low-temperature regime wherein transport is described via a co-tunneling mechanism is derived. 
This formula predicts the power-law temperature dependence of thermal conductance $\propto T^{2s+1}$ for a thermal environment of spectral density with the exponent $s$.
An accurate numerical simulation is performed using the quantum Monte Carlo method, and these predictions are confirmed for arbitrary thermal baths.
Our numerical calculation classifies the transport mechanism, and shows that the noninteracting-blip approximation quantitatively describes thermal conductance in the incoherent transport regime.
\end{abstract}

\section{Introduction}

Heat transport via small systems has recently attracted considerable attention because a lot of intriguing phenomena can emerge reflected from the properties of a system and the surrounding environment. For instance, quantized thermal conductances have been observed
in heat transport by phonons~\cite{Rego1998,Schwab2000} and photons~\cite{Meschke2006} in a manner similar to electric transport~\cite{Wees1988}.
Thermal rectification~\cite{Segal2005,Ruokola2009} and thermal transistors~\cite{Ojanen2008} have also been theoretically proposed in analogy to electronic devices.
Heat transport via a quasi one-dimensional material, e.g., carbon nanotubes, shows neither diffusive nor ballistic transport, which is currently categorized as anomalous transport~\cite{Lepri2016}.
Heat transport due to magnetic excitation 
is now a key ingredient in the field of spintronics~\cite{Adachi2013}.
Studying the general properties of thermal transport using typical systems is clearly an important subject not only for theoretical development but also for future experiments.

The spin-boson system is one of most common and important systems for describing a local discrete-level system embedded in a bosonic thermal environment~\cite{Weiss1999,Leggett1987}. 
This system has numerous applications, e.g., it is used to describe molecular junctions~\cite{Nitzan2006}, superconducting circuits~\cite{Makhlin2001}, and photonic waveguides with local two-level systems~\cite{LeHur2012}. Hence, it is regarded as a minimal model for describing a zero-dimensional object with discrete quantum levels surrounded by a bosonic environment.
One of the important problems here is to clarify the dissipative dynamics of the system near the equilibrium situation~\cite{Weiss1999}.
Depending on the properties of the thermal environment, the behavior of the autocorrelation function of the system changes from coherent oscillation to incoherent decay as a function of time. 
Intriguingly, at zero temperature, a quantum phase transition occurs when the coupling strength between the system and the environment is changed~\cite{Bray1982,Chakravarty1982}. 
The sub-ohmic environment induces a second-order phase transition~\cite{Bulla2003,Vojta2005,Winter2009,Vojta2009,Vojta2012,Chin2011}, while the ohmic case shows a Kostelitz-Thouless-type phase transition~\cite{Leggett1987,Anderson1971,Kosterlitz1976}. 
The super-ohmic case does not have a distinct phase transition but exhibits a crossover. 
In addition, the ohmic environment induces the Kondo effect~\cite{Hewson1997} at sufficiently low temperatures~\cite{Weiss1999,Leggett1987,Guinea1985a,Guinea1985b}. From this background in an equilibrium situation, it is quite natural to ask what happens if one considers heat transport in this system. 
Herein, we present systematic studies of heat transport via the spin-boson system and derive some exact results for this case.

A number of studies have investigated heat transport via spin-boson systems~\cite{Segal2010,Ruokola2011,Saito2013,Ren2010,Chen2013,Segal2014,Yang2014,Wang2015,Taylor2015}. 
Segal et al. introduced an iterative path-integral technique for numerical calculations to investigate the far-from-equilibrium regime~\cite{Segal2010}. 
Ruokola and Ojanen studied low-temperature properties using a perturbation method and discussed co-tunneling mechanisms~\cite{Ruokola2011}.
However, their methods do not seem to succeed in reproducing low-temperature properties, e.g., the Kondo effect. 
Two of the present authors (TK and KS) have focused on the transport properties in an ohmic environment and found several Kondo signatures~\cite{Saito2013}, including the $T^3$-temperature dependence of the thermal conductance. Herein, we advance in this direction and cover arbitrary types of environments. 
We consider a general picture for understanding the transport properties at extremely low temperatures for the whole regime of spectral densities and quantitatively characterize the transport mechanism for all temperature regimes.

We present our findings in this paper to distinguish them from existing literature.  
First, we derived an {\it asymptotically exact} expression for the thermal conductance in the extremely low-temperature regime, reproducing the aforementioned $T^3$-temperature dependence of thermal conductance in the ohmic case. 
Our formula is asymptotically exact in the co-tunneling transport regime and predicts power-law temperature dependences $\propto T^{2s+1}$ for the thermal environment of spectral density with the exponent $s$.
Second, we performed accurate numerical calculations to investigate thermal conductance over the entire temperature regime. We confirmed the temperature dependencies predicted by our expressions for the co-tunneling and the sequential tunneling transport regimes. 
Furthermore, we found that the noninteracting-blip approximation (NIBA)~\cite{Weiss1999} describes thermal conductance in the incoherent tunneling regime {\it accurately}. 
In table~\ref{table:tunneling_process} the transport mechanisms for each regime are summarized and relevant analytical descriptions are presented. 
In the table, sequential tunneling, co-tunneling, and NIBA imply the analytical descriptions based on the approximate form [equation (\ref{sequential_conductance})], the asymptotically exact expression [equation~(\ref{eq:cotunneling})], and the analytical descriptions based on the NIBA expression[equation~(\ref{exact_kappa2})] with equations~(\ref{eq:NIBAformula}) and (\ref{self-energy}).

\begin{table}
\caption{\label{table:tunneling_process}Summary of the relevant transport process. Here, $\Delta_{\rm eff}$ is an effective tunneling amplitude [see equations (\ref{eq:OhmicDeltaEff}) and (\ref{delta_eff_super_ohmic})] and $T^*$ is the crossover temperature [see equation~(\ref{eq:CrossoverTemperature})].
The last column shows the temperature dependences of the thermal conductance, where ``Schottky'' indicates a Schottky-type temperature dependence proportional to 
$e^{-\hbar \Delta_{\rm eff}/k_{\rm B}T}/T^2$.
The temperature dependence of NIBA is complex in general, and the symbol $(*)$ indicates the high-temperature limit.}
\begin{indented}
\item[]\begin{tabular}{@{}llll}
\br
Exponent & Condition & Transport process & Dependence \\
\mr
$0<s<1$ & $\alpha<\alpha_c$, $k_{\rm B} T \ll \hbar \Delta_{\rm eff}$ & Co-tunneling & $T^{2s+1}$ \\
(sub-ohmic) & $\alpha<\alpha_c$, $\hbar \Delta_{\rm eff}\lesssim k_{\rm B} T$ & Incoherent tunneling (NIBA) & \\
& $\alpha>\alpha_c$, arbitrary temperature & Incoherent tunneling (NIBA) & \\
\mr
$s=1$ & $\alpha<1$, $k_{\rm B} T \ll \hbar \Delta_{\rm eff}$ & Co-tunneling & $T^3$ \\
(ohmic) & $\alpha<1$, $\hbar \Delta_{\rm eff}\lesssim k_{\rm B} T$ & Incoherent tunneling (NIBA)& $T^{2\alpha - 1}(*)$ \\
& $\alpha>1$, arbitrary temperature & Incoherent tunneling (NIBA)  
& $T^{2\alpha - 1}$ \\
\mr
$1<s<2$ & $k_{\rm B} T \ll \hbar \Delta_{\rm eff}$ & Co-tunneling 
& $T^{2s+1}$ \\
(super-ohmic) & $\hbar \Delta_{\rm eff} \lesssim k_{\rm B}T < k_{\rm B} T^*$ & Sequential tunneling & Schottky \\
& $k_{\rm B} T^* < k_{\rm B} T$ & Incoherent tunneling (NIBA) &\\
\mr $s\ge2$&
$k_{\rm B} T \ll \hbar \Delta_{\rm eff}$ 
& Co-tunneling & $T^{2s+1}$ \\
(super-ohmic) & $\hbar \Delta_{\rm eff} \lesssim k_{\rm B}T$ & Sequential tunneling & Schottky \\
\br
\end{tabular}
\end{indented}
\end{table}

The paper is organized as follows.
In section 2, we introduce the model and explain the Meir-Wingreen-Landauer-type formula. 
In section 3, we classify the transport mechanism and derive an asymptotically exact expression that is valid in the co-tunneling transport regime.
We perform numerical calculation using the quantum Monte Carlo method, and compare the results with analytic approximations in section 4.
In section 5, we summarize our work. 

\section{Formulation}

\subsection{Model}

We consider heat transport via a local quantum system coupled to two reservoirs denoted by L and R.
The model Hamiltonian is given by
\begin{eqnarray}
& & H= H_{\rm S} + \sum_{\nu={\rm L},{\rm R}} H_{\nu} 
+ \sum_{\nu={\rm L},{\rm R}} H_{{\rm I},\nu}, \\
& & H_{\rm S} = \frac{p^2}{2m} + V(x), \\
& & H_{\nu} = \sum_{k} 
\left( \frac{p_{\nu k}^2}{2m_{\nu k}} + 
\frac12 m_{\nu k} \omega_{\nu k}^2 x_{\nu k}^2 \right), \\
& &  H_{\rm{I},\nu} = \sum_{k} \left(-C_{\nu k} x_{\nu k} x
+ \frac{C_{\nu k}^2}{2m_{\nu k} \omega_{\nu k}^2} x^2 \right),
\end{eqnarray}
where $H_{\rm S}$, $H_{\nu}$, and $H_{{\rm I},\nu}$ describe the local system, the reservoir $\nu$ ($={\rm L},{\rm R}$), and the interaction between them, respectively.
The operators $p$ and $x$ are the momentum and position for the local system, respectively, and $V(x)$ is the potential energy.
The reservoirs comprise multiple phonon (or photon) modes, which are described in general by harmonic oscillators with frequency $\omega_{\nu k}$ and mass $m_{\nu k}$, where the subscript denotes the phonon (photon) wavenumber $k$ in the reservoir $\nu$.
The momentum and position of an individual oscillator are denoted by $p_{\nu k}$ and $x_{\nu k}$, respectively.
For simplicity, the system-reservoir coupling $H_{\rm{I},\nu}$ is consider as a bilinear form of $x$ and $x_{\nu k}$, and the interaction strength is denoted by $C_{\nu k}$.
The second term of $H_{\rm{I},\nu}$ is a counter term to cancel the potential renormalization due to the reservoirs.

\begin{figure}[tb]
\begin{center}
\includegraphics[width=0.45\linewidth]{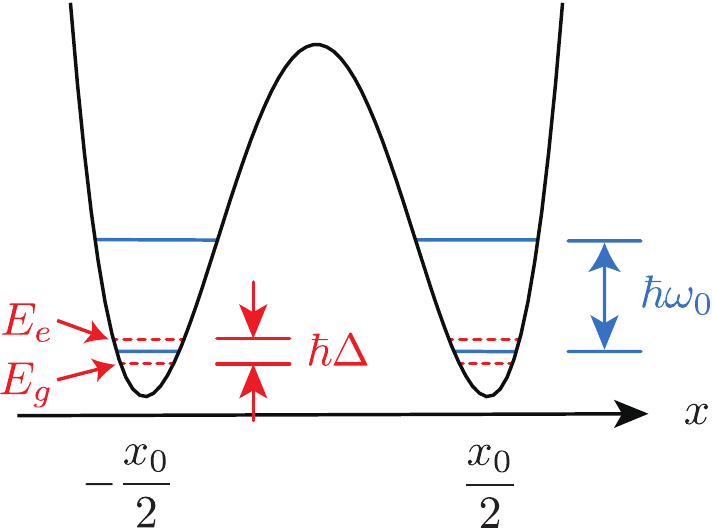}
\caption{
Symmetric double-well potential of the local system.
An energy spacing of quantum levels in each well is $\hbar\omega_0$ (indicated by the blue sold lines), and an energy splitting due to quantum tunneling  (indicated by the red dashed lines) is $\hbar\Delta = E_e - E_g$, where $E_g$ and  $E_e$ are the ground-state energy and the first-excited-state energy, respectively.}
\label{fig:system}
\end{center}
\end{figure}

In this study, the potential energy $V(x)$ of the local system is considered as a double-well potential as shown in figure~\ref{fig:system}. 
We assume that the barrier height of the double-well potential is sufficiently large in comparison with $\hbar \omega_0$, where $\omega_0$ is the frequency of a small oscillation at the potential minima $x=\pm x_0/2$.
Then, quantum tunneling between the two wells induces small energy splitting $\hbar \Delta$ ($\ll \hbar \omega_0$) between the ground-state energy $E_g$ and the first excited energy $E_e$.

After truncating the local system into two states by considering the two lowest energy eigenstates, we obtain the spin-boson model~\cite{Leggett1987,Weiss1999}:
\begin{eqnarray}
	\label{Hamiltonian}
	& & H= H_{\rm S} + 
    \sum_{\nu={\rm L},{\rm R}} H_{\nu} 
    + \sum_{\nu={\rm L},{\rm R}} H_{{\rm I},\nu}, \\
    & & H_{\rm S}= -\frac{\hbar \Delta}{2} \sigma_x
    -\varepsilon \sigma_z, \\
	& & H_{\nu} = \sum_{k} 
    \hbar \omega_{\nu k } b_{\nu k}^{\dagger} b_{\nu k}, \\
	& & H_{{\rm I},\nu} = -
    \frac{\sigma_z}{2} \sum_{k} 
    \hbar \lambda_{\nu k}(b_{\nu k}+b_{\nu k}^{\dagger}).
\end{eqnarray}
Here, $\sigma_{i}$ $(i=x,y,z)$ is the Pauli matrix, $b_{\nu k}$ is an annihilation operator defined by
\begin{eqnarray}
b_{\nu k} = \sqrt{\frac{m_{\nu k}\omega_{\nu k}}{2\hbar}} \left( x_{\nu k} + \frac{i p_{\nu k}}{m_{\nu k} \omega_{\nu k}} \right), 
\end{eqnarray}
and $\lambda_{\nu k} = x_0 C_{\nu k}/\sqrt{2\hbar m_{\nu k} \omega_{\nu k}}$.
In the present model, we assign the localized states at the left (right) well as $\Ket{\downarrow}$ ($\Ket{\uparrow}$).
Throughout this study, we examine the symmetric double-well potential ($\varepsilon = 0$) and only use the bias term $\varepsilon \sigma_z$ to define the static susceptibility
\begin{eqnarray}
\chi_0 = \lim_{\varepsilon \rightarrow 0} \frac{\langle \sigma_z \rangle}{\varepsilon},
\label{eq:StaticSusceptibilityDefinetion}
\end{eqnarray}
where $\Braket{\cdots}$ implies an equilibrium average.
For the symmetric case ($\varepsilon = 0$), the system Hamiltonian $H_{\rm S}$ describes the tunneling splitting $\hbar \Delta$ between the ground state ($\sigma_x=+1$) and the first excited state ($\sigma_x=-1$).

The properties of the reservoirs are characterized by the spectral function
\begin{equation}
	I_{\nu}(\omega) \equiv \sum_{k} 
    \lambda_{\nu k}^2\delta(\omega-\omega_{\nu k}) ,
\end{equation}
which is considered to be continuous assuming that the number of phonon (photon) modes is large.
For simplicity, we assume the following simple for the spectral function~\cite{Leggett1987,Weiss1999}:
\begin{eqnarray}
	& & I_{\nu}(\omega) = \alpha_\nu \tilde{I}(\omega), \\
   & & \tilde{I}(\omega) = 2\omega \left(\frac{\omega}{\omega_c}\right)^{s-1} 
   e^{-\omega/\omega_c}, \label{eq:spect} 
\end{eqnarray}
where $\alpha_{\nu}$ is the dimensionless coupling strength between the two-state system and the reservoir $\nu$. 
To cut off high-frequency excitation, we introduced the exponential cutoff function $e^{-\omega/\omega_c}$, where $\omega_c$ is the cutoff frequency, which is considerably larger than other characteristic frequencies, e.g., $\Delta$, $\varepsilon/\hbar$, and $k_{\rm B}T/\hbar$.
The exponent $s$ in equation (\ref{eq:spect}) is crucial for determining the properties of the reservoirs.
The case $s=1$ is called ``ohmic,'' whereas the cases $s>1$ and $s<1$ are called ``super-ohmic" and ``sub-ohmic," respectively.

\subsection{Thermal conductance}

The heat current flowing from reservoir $\nu$ into the local two-state system is defined as follows:
\begin{eqnarray}
	J_\nu\equiv -\frac{dH_{\nu}}{dt} 
    = \frac{i}{\hbar}[H_{\nu},H] = -i\frac{\sigma_z}{2} \sum_{k} 
    \lambda_{\nu k} \hbar \omega_{\nu k} (-b_{\nu k}+b_{\nu k}^{\dagger}) .
	\label{current_opperator}
\end{eqnarray}
Using the standard technique of the Keldysh formalism~\cite{Rammer1984,Jauho1994,Jauho2007}, one can derive the Meir-Wingreen-Landauer-type formula~\cite{Meir1992} for the nonequilibrium steady-state heat current $\average{J_L}= -\average{J_R} \equiv \average{J}$ as follows~\cite{Ojanen2008,Ruokola2009,Saito2013,Saito2008}:
\begin{eqnarray}
	\average{J} =\frac{\alpha\gamma}{8}\int_0^{\infty}d(\hbar \omega) \,
		\hbar \omega \, \mathrm{Im}[\chi(\omega)]
		\tilde{I }(\omega)\left[n_L(\hbar \omega)-n_R(\hbar \omega)\right],
		\label{eq:exact_current}
\end{eqnarray}
where $\alpha=\alpha_L+\alpha_R$, $\gamma=4\alpha_L\alpha_R/\alpha^2$ is an asymmetric factor, $n_\nu(\omega)$ is the Bose distribution function in reservoir $\nu$, and $\chi(\omega)$ is the dynamical susceptibility of the two-state system defined by
\begin{eqnarray}
& & \chi(\omega) = - \frac{i}{\hbar} \int_0^{\infty} dt \, \langle [\sigma_z(t),\sigma_z(0)] \rangle e^{i\omega t}.
\end{eqnarray}
Equation~(\ref{eq:exact_current}) is derived in \ref{app:MierWingreeen}.
The linear thermal conductance is defined as
\begin{eqnarray}
	\kappa\equiv \lim_{\Delta T \rightarrow 0} \frac{\average{J}}{\Delta T}.
	\label{den_kappa}
\end{eqnarray}
Using the exact formula [equation (\ref{eq:exact_current})], the linear thermal conductance is given as
\begin{eqnarray}
	\kappa=\frac{\alpha\gamma k_{\rm B}}{8}\int_{0}^{\infty}d(\hbar \omega) \, \mathrm{Im}
	[\chi(\omega)]\tilde{I}(\omega)
    \left[\frac{\hbar \beta\omega/2}{\mathrm{sinh}(\hbar \beta\omega/2)}\right]^2 ,
	\label{exact_kappa}
\end{eqnarray}
where $\chi(\omega)$ is evaluated for the thermal equilibrium and $\beta = 1/(k_{\rm B}T)$.
Thus, we need to calculate the dynamical susceptibility $\chi(\omega)$ for evaluating the linear thermal conductance.

For convenience of discussion, we also introduce a symmetrized correlation function and its Fourier transformation:
\begin{eqnarray}
& & S(t) =\frac{1}{2}\Braket{\sigma_z(t)\sigma_z(0)+\sigma_z(0)\sigma_z(t)} ,\\
& & S(\omega) = \int_{-\infty}^{\infty} dt \, S(t) e^{i \omega t}.
\end{eqnarray}
From the fluctuation-dissipation theorem~\cite{Weiss1999}, the imaginary part of the dynamical susceptibility is related to $S(\omega)$ as
\begin{eqnarray}
& & S(\omega) = \hbar \, {\rm coth} \left(\frac{\hbar \beta \omega}{2} \right)
 {\rm Im}[\chi(\omega)] .
\end{eqnarray}
The thermal conductance is then rewritten using the correlation function $S(\omega)$ as
\begin{eqnarray}
	\kappa=\frac{\alpha\gamma k_{\rm B}}{8}\int_{0}^{\infty}d\omega \, {\rm tanh} \left(\frac{\hbar \beta \omega}{2} \right) 
    S(\omega) \tilde{I}(\omega)
    \left[\frac{\hbar \beta\omega/2}{\mathrm{sinh}(\hbar \beta\omega/2)}\right]^2 .
	\label{exact_kappa2}
\end{eqnarray}

\section{Classification of Transport Processes}
\label{sec:Classification}

The dynamics of dissipative two-state systems have long been studied using a number of approximations~\cite{Leggett1987,Weiss1999}.
In this section, we re-examine such analytic approximations from the viewpoint of heat transport.
In section~\ref{sec:EffectiveTunnelingAmplitude}, we first consider the effective tunneling amplitude and discuss a quantum phase transition driven by strong system-reservoir coupling.  
Next, we consider the three mechanisms, which we call ``sequential tunneling'' (section~\ref{sec:SequentialTunneling}), ``co-tunneling'' (section~\ref{sec:Cotunneling}), and ``incoherent tunneling'' (section~\ref{sec:IncoherentTunneling}) following in
the previous literatures~\cite{Leggett1987,Weiss1999,Ruokola2011,Agarwalla2017}. We derive analytic expressions for the thermal conductance in each transport process.
We also introduce NIBA in section~\ref{sec:NIBA}.

In this section, we show two novel results of our study.
The first concerns the co-tunneling process. 
We derive an asymptotically exact formula for the co-tunneling process by utilizing the generalized Shiba relation.
This formula always holds at low temperatures for an arbitrary exponent ($s$) as long as the ground state of the system is delocalized.
The second result is related to the incoherent tunneling.
In particular, we find that the Markov approximation is inadequate to describe the thermal conductance in the incoherent tunneling regime.
Instead, the thermal conductance in this regime is well described by NIBA, which considers the non-Markovian properties of stochastic dynamics.
We show that NIBA quantitatively explains numerical calculations in section~\ref{sec:Result}.

\subsection{Effective tunneling amplitude and quantum phase transition}
\label{sec:EffectiveTunnelingAmplitude}

One important effect of the system-reservoir coupling is renormalization of the tunneling amplitude $\Delta$.
In this subsection, we briefly show the effective tunneling amplitude results obtained via adiabatic renormalization~\cite{Leggett1987,Weiss1999}.
A detailed derivation is given in \ref{app:AdiabaticRenormalization}.

\begin{figure}[tb]
	\centering
	\includegraphics[width=85mm]{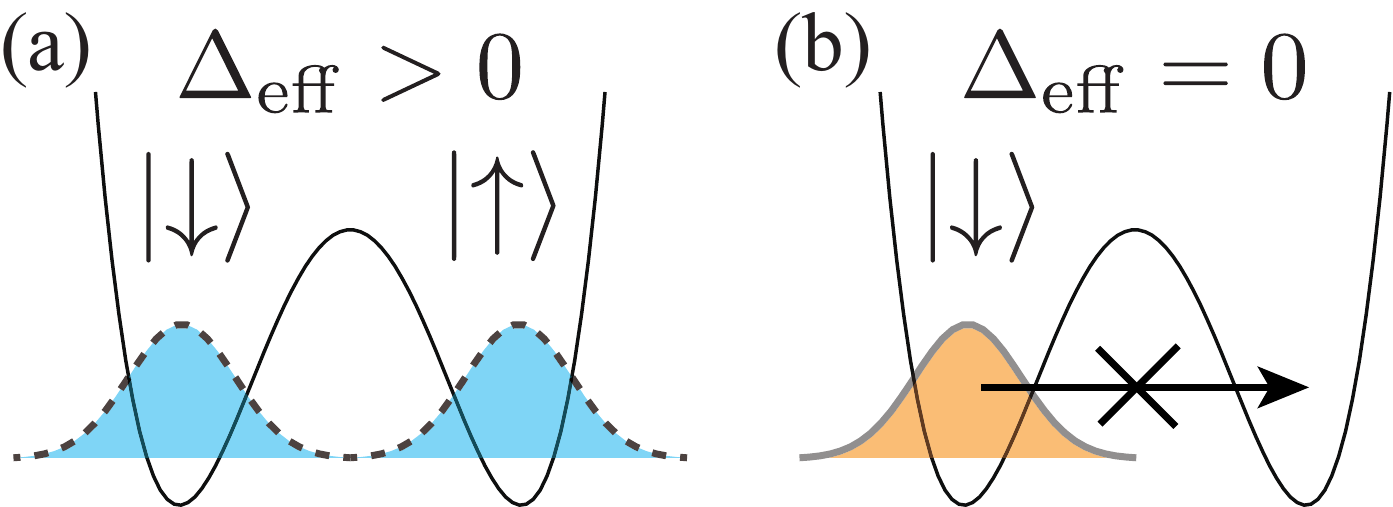}
	\caption{Schematics of the ground-state wavefunction (a) below the transition ($0 \le \alpha < \alpha_c$) and (b) above the transition ($\alpha_c < \alpha$).
    The former state is delocalized, whereas the latter is localized at one of the two wells.
   For the localized state, quantum tunneling between the two wells is forbidden since the overlap
   integral between the states in the two wells vanishes.}
    \label{fig:groundstate}
\end{figure}

For the ohmic case ($s=1$), the effective tunneling amplitude is given by
\begin{eqnarray}
\Delta_{\mathrm{eff}}=\left\{ \begin{array}{ll}
\displaystyle{[\Gamma(1-2\alpha)\cos(\pi\alpha)]^{1/2(1-\alpha)}
\Delta \left(\frac{\Delta}{\omega_c}\right)^{\alpha/(1-\alpha)}}, & (0\le \alpha<1), \\
0, & (1 < \alpha ).
\end{array} \right. 
\label{eq:OhmicDeltaEff}
\end{eqnarray}
This result indicates a phase transition at zero temperature, for which the critical value of the system-reservoir coupling is $\alpha=1$~\cite{Chakravarty1982,Bray1982}.
For system-reservoir couplings below the transition ($0\le \alpha<1$), the ground state is non-degenerate, as shown in figure~\ref{fig:groundstate}~(a), indicating the coherent superposition of the two localized states $\Ket{\uparrow}$ and $\Ket{\downarrow}$.
We call this ground state ``delocalized.''
For strong system-reservoir couplings above the transition ($\alpha > 1$), the coherent superposition of the two localized states is completely broken, leading to the doubly-degenerate ground states shown in figure~\ref{fig:groundstate}~(b).
We call this ground state ``localized.''
In this localized regime, quantum tunneling between the wells is forbidden at zero temperature since there is no mixing ($\Delta_{\rm eff}=0$) between the two localized states.
Thus, the present quantum phase transition can be recognized as a ``localization'' transition that separates the delocalized and localized regimes at zero temperature.

For the sub-ohmic case ($s<1$), the adiabatic renormalization always leads to an effective tunneling amplitude of zero ($\Delta_{\rm eff} = 0$).
This is correct in the limit $\Delta/\omega_c\rightarrow0$, as discussed in a previous study~\cite{Leggett1987}.
However, for a finite value of $\Delta/\omega_c$, the naive adiabatic renormalization procedure yields incorrect results and should be improved.
In subsequent theoretical studies~\cite{Kehrein1996_1,Kehrein1996_2}, it was found that the localization transition actually occurred at a critical system-reservoir coupling ($\alpha = \alpha_c$), where the critical value $\alpha_{\rm c}$ depended on both $s$ and $\Delta/\omega_c$.
The existence of the localization transition was also confirmed via numerical calculations~\cite{Vojta2005,Winter2009}.
In summary, for the sub-ohmic case, the ground state is delocalized for $0 \le \alpha < \alpha_c$, as shown in figure~\ref{fig:groundstate}~(a), and localized for $\alpha_c <\alpha$, as shown in figure~\ref{fig:groundstate}~(b).

For the super-ohmic case ($s>1$), the effective tunneling amplitude is always finite:
\begin{eqnarray}
\label{delta_eff_super_ohmic}
\Delta_\mathrm{eff} &=&\Delta \exp \left(-\alpha\Gamma(s-1)\right),
\end{eqnarray}
where $\Gamma(z)$ is the Gamma function.
Therefore, there is no localization transition and the ground state is always delocalized, as shown in figure~\ref{fig:groundstate}~(a).

\subsection{Sequential tunneling}
\label{sec:SequentialTunneling}

\begin{figure}[tb]
	\centering
	\includegraphics[width=140mm]{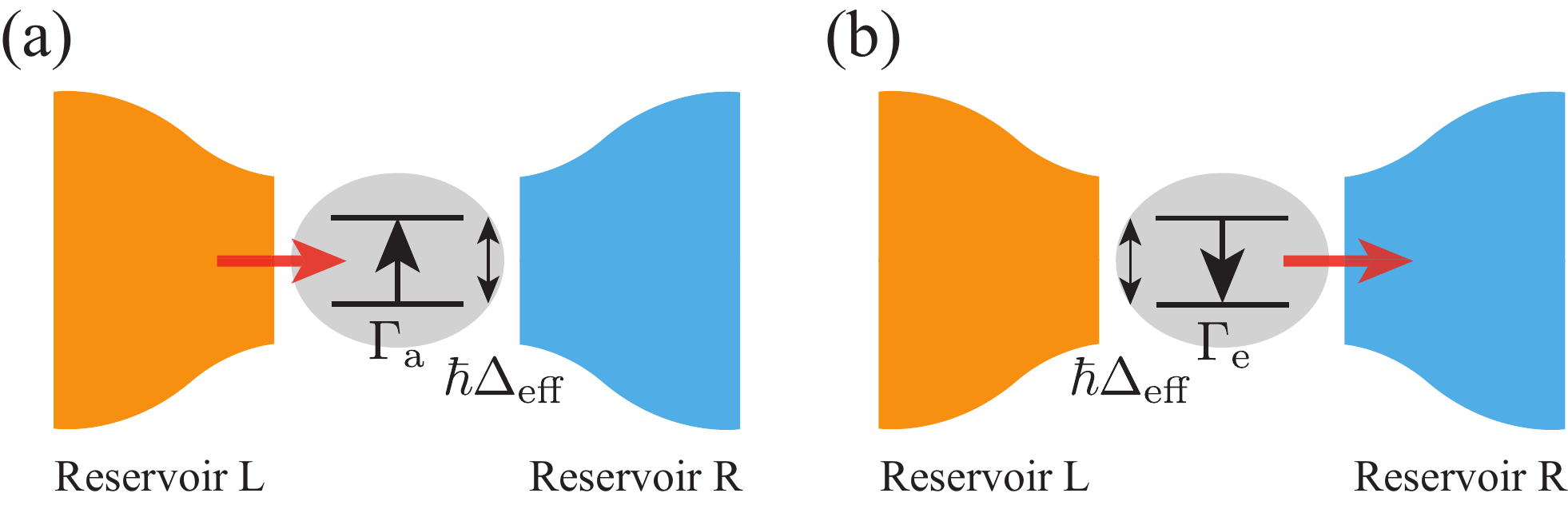}
	\caption{Schematic of the sequential tunneling process.  
    Heat transport occurs by a combination of (a) phonon (photon) absorption and (b) phonon (photon) emission.}
    \label{fig:sequential}
\end{figure}

For weak system-reservoir couplings ($\alpha \ll 1$), the system and the reservoirs are almost decoupled and the interaction Hamiltonian $H_{{\rm I},\nu}$ can be regarded as a perturbation.
For the second-order perturbation, the system dynamics are described by a stochastic transition between the ground state ($\sigma_x = +1$) and the excited state ($\sigma_x = -1$), as shown in figure~\ref{fig:sequential}.
The transition from the ground state to the excited state involves phonon (photon) absorption, and the inverse transition involves phonon (photon) emission.
A combination of these two processes induces heat transport.
We refer to this type of transport process as ``sequential tunneling'' by analogy with the electronic transport process through quantum dots.
The transition rates for the process of phonon (photon) absorption and emission are calculated based on Fermi's golden rule as follows~\cite{Leggett1987}:
\begin{eqnarray}
\Gamma_{\rm a} = \frac{\pi}{2}n_{\rm B}(\Delta) I(\Delta), \quad \quad
\Gamma_{\rm e} = \frac{\pi}{2}(n_{\rm B}(\Delta) + 1) I(\Delta), 
\end{eqnarray}
where $I(\omega) = I_{\rm L}(\omega) + I_{\rm R}(\omega)$ and $n_{\rm B}(\omega) = (e^{\hbar \beta \omega}-1)^{-1}$ is a Bose distribution function. 
Using these transition rates, the stochastic dynamics of the system are described using the Lindblad equation
\begin{eqnarray}
\frac{d\rho(t)}{dt} = - \frac{i}{\hbar} [ H_{\rm S}, \rho(t)] + \sum_{j={\rm e},{\rm a}}
\Gamma_j \left( L_j \rho(t) L_j^{\dagger}
- \frac{1}{2}(L_j^{\dagger} L_j \rho +
\rho L_j^{\dagger} L_j ) \right),
\end{eqnarray}
where $\rho(t)$ is a density matrix of the system, $L_{\rm e} = \sigma_x^+ \equiv(\sigma_z-i\sigma_y)/2$, and $L_{\rm a} = \sigma_x^- \equiv(\sigma_z+i\sigma_y)/2$.
By solving this equation, we obtain the symmetrized correlation function as
\begin{eqnarray}
& & S(\omega) = \frac{4\Gamma(\Delta^2 + \Gamma^2)}{[(\omega-\Delta)^2 + \Gamma^2][(\omega+\Delta)^2 + \Gamma^2]}, 
\end{eqnarray}
where $\Gamma = (\Gamma_{\rm e} + \Gamma_{\rm a})/2$.
The correlation function $S(\omega)$ has two peaks at $\omega = \pm \Delta$, reflecting the coherent system dynamics.
Because $\Gamma \ll \Delta$ always holds in the weak-coupling regime, the correlation function is approximated as 
\begin{eqnarray}
S(\omega) \simeq \pi [\delta(\omega - \Delta) + \delta(\omega + \Delta)],
\label{eq:SomegaSequential}
\end{eqnarray}
where $\delta(x)$ is the delta function.
The thermal conductance for the weak coupling regime is obtained by substituting equation~(\ref{eq:SomegaSequential}) into equation~(\ref{exact_kappa2}) as follows:
\begin{eqnarray}
\kappa\simeq \frac{\pi \alpha \gamma k_B}{8} 
{\rm tanh}\left(\frac{\hbar\beta\Delta}{2}\right) \tilde{I}(\Delta) 
\left[\frac{\hbar \beta\Delta/2}{\mathrm{sinh}{(\hbar \beta\Delta/2})}\right]^{2}.
\label{sequential_conductance}
\end{eqnarray}
This result is identical to the formula derived in previous research~\cite{Segal2005} and \cite{Saito2013} using the master equation approach and is consistent with the perturbation theory~\cite{Ruokola2011}.
For actual comparison with the numerical simulation in section~\ref{sec:Result}, we improve the approximation by replacing $\Delta$ with $\Delta_{\rm eff}$ using adiabatic renormalization (see section~\ref{sec:EffectiveTunnelingAmplitude}).

The formula for sequential tunneling [equation~(\ref{sequential_conductance})] is valid when
\begin{eqnarray}
\Gamma = \frac{\pi}{4} (2n_{\rm B}(\Delta_{\rm eff}) + 1) I(\Delta_{\rm eff}) \ll \Delta_{\rm eff}.
\end{eqnarray}
For the sub-ohmic case ($s<1$), this condition is never satisfied, indicating the absence of a sequential tunneling regime.
For the ohmic case ($s=1$), the condition is equivalent to $\alpha \ll 1$, whereas for the super-ohmic case ($s>1$), the condition is always satisfied for a moderate temperature ($k_{\rm B} T \sim \hbar \Delta_{\rm eff}$). 
At high temperatures ($k_{\rm B} T \gg \hbar \Delta_{\rm eff}$), the condition is always satisfied for $s\ge 2$, whereas for $1<s<2$, it becomes
\begin{eqnarray}
T < T^* = \frac{\hbar \omega_c}{\alpha k_{\rm B}} \left(\frac{\Delta_{\rm eff}}{\omega_c}\right)^{2-s},
\label{eq:CrossoverTemperature}
\end{eqnarray}
where $T^*$ is the crossover temperature.

The formula for sequential tunneling [equation~(\ref{sequential_conductance})] predicts the exponential decrease in the thermal conductance as the temperature is lowered. At low temperatures, the thermal conductance behaves as $\kappa \propto e^{-\hbar\Delta_\mathrm{eff}/k_{\rm B}T}/T^2$; this is because the transition from the ground state to the excited state is strongly suppressed if the thermal fluctuation is smaller than the effective energy splitting, i.e., when $k_{\rm B}T \ll \hbar \Delta_{\rm eff}$.
When the sequential tunneling process is strongly suppressed at low temperatures, equation (\ref{sequential_conductance}) becomes invalid since another process becomes dominant, as discussed in the next subsection.

\subsection{Co-tunneling and an asymptotically exact formula}
\label{sec:Cotunneling}

\begin{figure}[tb]
	\centering
	\includegraphics[width=70mm]{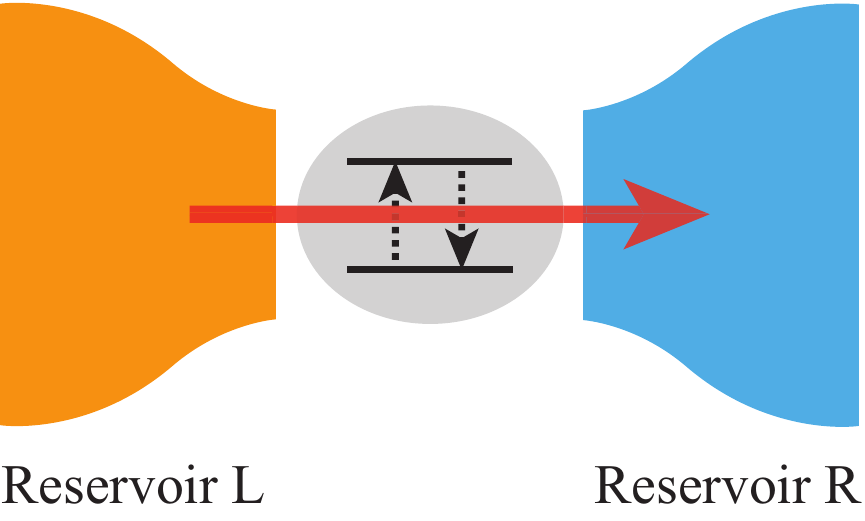}
	\caption{Schematic of the co-tunneling process. At $k_{\rm B}T\ll\hbar\Delta_\mathrm{eff}$, heat transport via a virtual excitation in the local system is dominant.}
    \label{fig:cotunneling}
\end{figure}

At low temperatures, heat transport via the virtual excitation of the local two-state system becomes dominant (see figure~\ref{fig:cotunneling});
this transport process is known as ``co-tunneling'' by analogy with the electronic transport process through quantum dots.
In a previous study~\cite{Ruokola2011}, an analytical expression for thermal conductance was derived using the fourth-order perturbation theory with respect to the interaction $H_{{\rm I},\nu}$.
However, in this calculation the renormalization of the tunneling amplitude at a low temperature has not been considered.

Here, we derive a new asymptotically exact formula for the thermal conductance without any approximations.
For this purpose, we focus on an asymptotically exact relation called the generalized Shiba relation~\cite{Shiba1975,Sassetti1990}:
\begin{eqnarray}
\lim_{\omega \to 0+} \frac{S(\omega)}{\tilde{I}(\omega)} =\pi\alpha\left(\frac{\hbar \chi_0}{2}\right)^{2},
\end{eqnarray}
where $\chi_0$ is the static susceptibility defined in equation (\ref{eq:StaticSusceptibilityDefinetion}).
This exact relation holds at low temperatures ($k_{\rm B} T \ll \hbar \Delta_{\rm eff}$) for arbitrary environments and arbitrary system-reservoir couplings.
At low temperatures ($k_{\rm B} T \ll \hbar \Delta_{\rm eff}$), the dominant contribution to the integral of equation~(\ref{exact_kappa}) comes from the low-frequency part ($0 \le \hbar \omega \simeq k_{\rm B} T \ll \hbar \Delta_{\rm eff}$) due to the factor of the Bose distribution function.
By substituting the low-frequency asymptotic form $S(\omega) \simeq \pi\alpha (\hbar \chi_0/2)^2 \tilde{I}(\omega)$ into equation (\ref{exact_kappa}), we obtain
\begin{eqnarray}
\label{eq:cotunneling}
\kappa\simeq \frac{\pi k_{\rm B} (\hbar \chi_0)^2}{8}
\int_0^{\infty} d\omega \, I_{\rm L}(\omega) I_{\rm R}(\omega)
\left[ \frac{\hbar \beta \omega/2}{\sinh(\hbar \beta \omega/2)}\right]^2.
\end{eqnarray}
This expression is similar to the co-tunneling formula in previous studies~\cite{Ruokola2011,Wu2011,Agarwalla2017} but significantly differs in terms of static susceptibility, $\chi_0$, which considers higher-order processes.
Equation~(\ref{eq:cotunneling}) can be rewritten as
\begin{eqnarray}
\label{eq:cotunneling2}
& & \kappa\simeq \frac12 \pi k_B \alpha_{\rm L} \alpha_{\rm R} \omega_c^{3} ( \hbar \chi_0)^2\left(\frac{k_{\rm B} T}{\hbar \omega_{\rm c}}\right)^{2s+1} \! F(s), \\
& & F(s)=\int_{0}^{\infty}dx \, x^{2s} \left[\frac{x/2}{\sinh({x/2})}\right]^{2},
\label{eq:FsDefinition}
\end{eqnarray}
where $F(s)$ is a dimensionless function of $s$.
Thus, we find that the thermal conductance $\kappa$ is proportional to $T^{2s+1}$ at low temperatures.
The same temperature dependence has been derived by the perturbation theory~\cite{Ruokola2011,Wu2011,Agarwalla2017}.
However, the perturbation theory cannot treat renormalization effect due to higher-order processes on the static susceptibility, and fails in predicting a correct prefactor including $\chi_0$.
In contrast, the present result given in equation~(\ref{eq:cotunneling}) is asymptotically exact, incorporating the renormalization effect appropriately.

The co-tunneling formula [equation (\ref{eq:cotunneling})], a new formula that is first derived in the present study, holds universally at low temperatures for an arbitrary exponent, $s$, as long as the ground state of the system is delocalized ($\Delta_{\rm eff} > 0$)
In a previous study~\cite{Saito2013}, the thermal conductance in the ohmic case ($s=1$) was shown to be proportional to $T^3$, which is consistent with equation (\ref{eq:cotunneling}), and this $T^3$-dependence was discussed in terms of the emergence of the Kondo effect. However, it is worth nothing that the power-law temperature dependences are derived in an unified way even in non-ohmic cases. These temperature dependences result from nontrivial many-body effects due to strong mixing between the system and the reservoirs.

\subsection{Incoherent tunneling: the Markov approximation}
\label{sec:IncoherentTunneling}

\begin{figure}[tb]
	\centering
	\includegraphics[width=70mm]{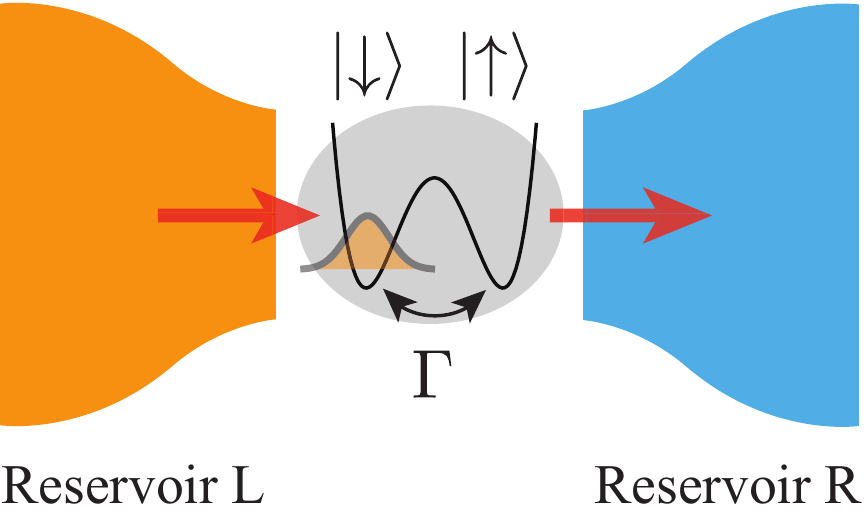}
	\caption{Schematic of the incoherent tunneling process. 
    The wavefunction is localized in the two wells, and a stochastic
    transition occurs between them.}
    \label{fig:incoherent}
\end{figure}

For a strong reservoir-system coupling, the coherent superposition of the two localized states is completely broken.
In such a situation, heat transport is induced by stochastic dynamics between the two localized states $\Ket{\uparrow}$ and $\Ket{\downarrow}$, as shown in figure~\ref{fig:incoherent}.
We call this transport process ``incoherent tunneling.''

Within the Markov approximation~\cite{Fisher1985,Grabert1985,Weiss1985}, the stochastic dynamics of the system are described by the master equation
\begin{eqnarray}
\label{eq:master_eq}
\frac{dP_{\rm L}(t)}{dt}=-\Gamma P_{\rm L}(t) + \Gamma P_{\rm R}(t),
\quad
\frac{dP_{\rm R}(t)}{dt}= \Gamma P_{\rm L}(t) - \Gamma P_{\rm R}(t),
\label{eq:masterEq}
\end{eqnarray}
where $P_{\rm L}(t)$ and $P_{\rm R}(t)$ ($=1- P_{\rm L}(t)$) are the probabilities that the wavefunctions of the system are localized at the well on the left-hand side ($\sigma_z = -1$) and that on the right-hand side ($\sigma_z = 1$), respectively, at time $t$.
The transition rate $\Gamma$ is calculated via second-order perturbation with respect to the Hamiltonian $H_{\rm S}$ as follows~\cite{Leggett1987}:
\begin{eqnarray}
\Gamma=\frac{\Delta^2}{2} \int_{-\infty}^\infty d\tau\,
e^{-Q_1(\tau)} \cos[Q_2(\tau)],
\label{eq:GammaDefinition} \\
Q_1(\tau)=\int_0^\infty d\omega\, \frac{I(\omega)}{\omega^2}
\coth\left(\frac{\hbar \beta\omega}{2}\right)[1-\cos(\omega\tau)],
\label{eq:Q1} \\
Q_2(\tau)=\int_0^\infty d\omega\, \frac{I(\omega)}{\omega^2}
\sin(\omega\tau).
\label{eq:Q2}
\end{eqnarray}
Note that this expression for the transition rate of incoherent tunneling is valid when $\hbar \Gamma \ll k_{\rm B}T$~\cite{Weiss1985}.
By solving the master equation [equation (\ref{eq:masterEq})], the symmetrized correlation function is calculated as
\begin{eqnarray}
S(\omega) = \frac{4\Gamma}{\omega^2 + 4\Gamma^2} \ .
\label{eq:SomegaIncoherent}
\end{eqnarray}
In contrast to sequential tunneling, $S(\omega)$ has only one peak at $\omega = 0$ with a width of $2\Gamma$, indicating the destruction of the superposition of the two localized states.

The long-term dynamics are well described by the Markov approximation~\cite{Leggett1987}.
Therefore, one may expect that the thermal conductance in the incoherent tunneling regime would be well approximated by substituting equations (\ref{eq:GammaDefinition})-(\ref{eq:SomegaIncoherent}) into equation (\ref{exact_kappa2}).
However, the results of the Markov approximation show clear deviation from the numerical results, as discussed in section~\ref{sec:Result}.
The reason for this is summarized as follows.
Note that incoherent tunneling occurs when $\hbar \Gamma \ll k_{\rm B}T$.
Under this condition, the integrand of equation~(\ref{exact_kappa2}) is proportional to $\omega^{s-2}$ for $\Gamma \ll \omega \ll k_{\rm B}T/\hbar$ since $S(\omega) \propto \omega^{-2}$ [see equation~(\ref{eq:SomegaIncoherent})].
Then, the integral in equation (\ref{exact_kappa2}) diverges if the high-frequency cut-off occurring due to the Bose distribution function is absent.
This indicates that the high-frequency part of the integral in equation~(\ref{exact_kappa2}) makes the dominant contribution to the thermal conductance.
Although the Markov approximation yields reasonable results for the low-frequency behavior of $S(\omega)$, it fails to reproduce the accurate high-frequency behavior of $S(\omega)$ in general, leading to incorrect results for the thermal conductance.

\subsection{NIBA}
\label{sec:NIBA}

To study the short-term (high-frequency) dynamics in the incoherent tunneling regime, we introduce the NIBA, which is a natural extension of the Markov approximation in the previous subsection~\cite{Leggett1987,Liu2017}.
In NIBA, the symmetrized correlation function is calculated in a manner same as that followed in a previous study~\cite{Weiss1999}:
\begin{eqnarray}
S(\omega)=2\mathrm{Re}\left[\frac{1}{-i\omega
+ \Sigma(-i\omega)}\right], 
\label{eq:NIBAformula}
\end{eqnarray}
where $\Sigma(\lambda = -i\omega)$ is the frequency-dependent self-energy defined as
\begin{eqnarray}
\label{self-energy}
\Sigma(\lambda) = \Delta^2\int_0^\infty d\tau \, 
e^{-\lambda\tau} e^{-Q_1(\tau)}\cos[Q_2(\tau)].
\end{eqnarray}
Here, $Q_1(\tau)$ and $Q_2(\tau)$ are given by equations (\ref{eq:Q1}) and (\ref{eq:Q2}), respectively.
The thermal conductance is then calculated by substituting equations~(\ref{eq:NIBAformula}) and (\ref{self-energy}) into equation~(\ref{exact_kappa2}).
From the definition, it is easy to check that NIBA reproduces the Markov approximation if we neglect the frequency dependence of the self-energy and replace it with the zero-frequency value $\Sigma(0) = 2\Gamma$.
Since NIBA appropriately considers the non-Markovian properties, it is suitable to describe the thermal conductance in the incoherent tunneling regime.

The condition for NIBA is well known~\cite{Leggett1987,Weiss1999}.
As expected from the fact that NIBA is an extension of the Markov approximation, it works well for the incoherent tunneling regime.
Roughly, the incoherent tunneling mechanism becomes crucial in a regime wherein both the sequential tunneling formula and the co-tunneling formula fail. (a) NIBA holds at moderate-to-high temperatures in the sub-ohmic ($s<1$) and ohmic cases ($s=1$). (b) It holds for $T>T^*$ in the super-ohmic case of $1<s<2$, where $T^*$ is the crossover temperature discussed in section~\ref{sec:SequentialTunneling}.
Note that NIBA never holds for $s\ge 2$ since the crossover temperature $T^*$ diverges.

Here, the NIBA has been introduced to improve the Markov approximation in the incoherent regime.
This introduction of the NIBA may give impression to the readers that the NIBA is a good approximation only in the incoherent regime.
However, the NIBA is known to be applicable for a wider parameter region not restricted to the incoherent regime~\cite{Weiss1999}.
The NIBA holds also in the weak coupling regime ($\alpha\ll 1$) at arbitrary temperature for the unbiased case ($\varepsilon=0$), where the interblip interaction is shown to be much weaker than the the intrablip interaction (for detailed discussion, see Sec. 21.3 in Ref.~\cite{Weiss1999}).
For this reason, NIBA yields almost the same result as the sequential tunneling formula or the co-tunneling formula if the system-reservoir coupling is sufficiently weak.

In section~\ref{sec:Result}, we show that NIBA is an excellent approximation for reproducing the numerical results for a wide region of the parameter space at moderate-to-high temperatures. 
Thus, the short-term (high-frequency) non-Markovian behavior in the system dynamics is important for calculating the thermal conductance in the incoherent tunneling regime.
\section{Numerical Results and Comparison with Analytical Formulas}
\label{sec:Result}

While the analytical approaches discussed in the previous section are sufficiently powerful for clarifying the mechanism of heat transport in a two-state system, the detailed conditions justifying each approximation are not trivial.
To understand all features of heat transport, unbiased numerical simulation without any approximation would be helpful.
In this section, we therefore perform numerical simulations based on the quantum Monte Carlo method and compare the simulation results with the analytical formulas introduced in section~\ref{sec:Classification}.
After briefly describing the numerical method in section~\ref{sec:Numerics}, we separately consider the ohmic (section~\ref{sec:ResultOhmic}), sub-ohmic (section~\ref{sec:ResultSubOhmic}), and super-ohmic cases (sections~\ref{sec:ResultSuperOhmic} and \ref{sec:ResultSuperOhmic2}).

The dynamics of the spin-boson model has been studied by using various numerical methods~\cite{Velizhanin2008,Segal2013,Boudjada2014,Wong2008,Schroder2016,Ballestero2017,Volker1998}.
However no systematic comparisons between analytical approximations and numerical simulations has been performed in the context of heat transport near thermal equilibrium.
This comparison allows us to discuss the validity of various approximations critically.

\subsection{Numerical method}
\label{sec:Numerics}

For numerical simulations, we employ the continuous-time quantum Monte Carlo (CTQMC) algorithm proposed in a previous study~\cite{Winter2009}.
According to this algorithm, the partition function is rewritten in path-integral form with respect to an imaginary time path, $\sigma_z(\tau)$, and the weight of this path is defined.
Then, we apply the Monte Carlo method to this representation using the cluster update algorithm~\cite{Gubernatis2016}.
The details of the CTQMC method are given in \ref{app:CTQMC}.

Using the CTQMC method, we evaluate the imaginary time spin correlation function $C(\tau)$ and its Fourier transform as follows:
\begin{eqnarray}
C(\tau) = \average{\sigma_z(\tau)\sigma_z(0)}, \\
\label{fourier}
C(i\omega_n)=\int_{0}^{\hbar \beta}d\tau e^{i\omega_n \tau}C(\tau),
\label{eq:CMatsubara}
\end{eqnarray}
where $\sigma_z(\tau) = e^{\tau H/\hbar} \sigma_z e^{-\tau H/\hbar}$.
The dynamical susceptibility $\chi(\omega)$ is obtained from $C(i\omega_n)$ via analytical continuation as follows:
\begin{eqnarray}
\label{pade}
\chi(\omega)=C(i\omega_n\rightarrow\omega+i\delta).
\end{eqnarray}
Analytical continuation is performed by Pad\'e approximation~\cite{Baker1975,Vidberg1977} or by fitting the imaginary time spin correlation function's Fourier transform to the Lorentzian function~\cite{Volker1998}.
For details, see \ref{app:CTQMC}.

\subsection{The ohmic case ($s=1$)}
\label{sec:ResultOhmic}

\begin{figure}[tbp]
	\centering
	\includegraphics[height=10.0cm]{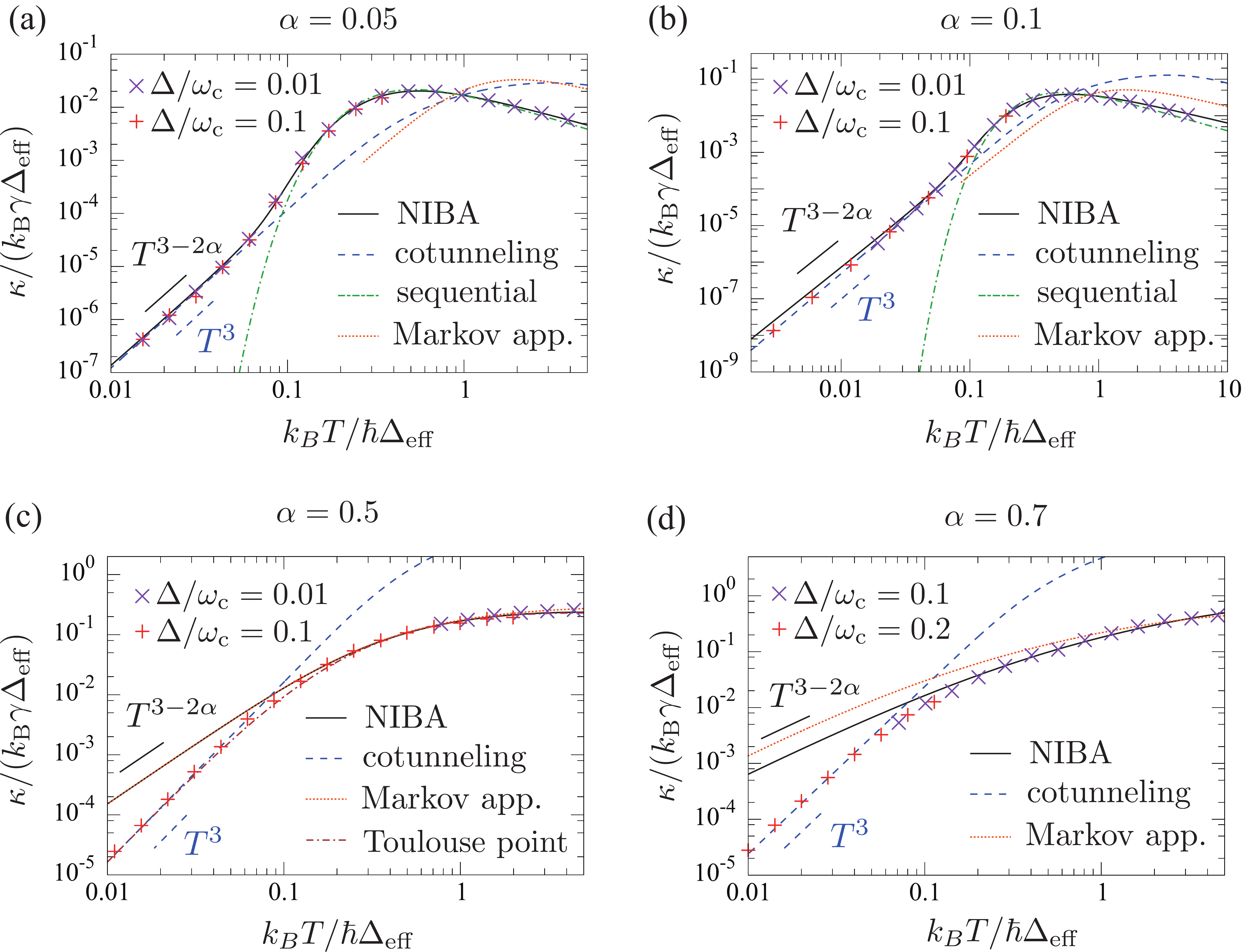}
	\caption{The temperature dependence of the thermal conductance for (a) $\alpha = 0.05$, (b) $0.1$, (c) $0.5$, and (d) $0.7$.
   The symbols indicate the numerical results obtained using the CTQMC method. 
  The black solid, green dot-dashed, blue dashed, and orange dotted lines represent NIBA, sequential tunneling formula, co-tunneling formula, and Markov approximation for incoherent tunneling, respectively. 
In (c), the exact solution for the Toulouse point ($\alpha = 0.5$) is indicated by the brown dotted line.}
	\label{fig:ohmic_conductance}
\end{figure}

In figure~\ref{fig:ohmic_conductance}, we show the thermal conductances for $\alpha = 0.05$, $0.1$, $0.5$, and $0.7$ as functions of temperature.
We plot the graph using the normalized temperature $k_{\rm B}T/\hbar \Delta_{\rm eff}$ and the normalized thermal conductance $\kappa/(k_{\rm B} \gamma \Delta_{\rm eff})$, where $\Delta_{\rm eff}$ is the effective tunneling amplitude defined in equation (\ref{eq:OhmicDeltaEff}).
As shown in figure~\ref{fig:ohmic_conductance}, the numerical results fall on a universal scaling curve at each value of $\alpha$ regardless of the ratio $\Delta/\omega_c$ ($\ll 1$) obtained via this normalization.
This universal behavior is characteristic of the Kondo-like effect~\cite{Saito2013}.
In figure~\ref{fig:ohmic_conductance}~(c), we also show the exact solution (the Toulouse point) for $\alpha = 0.5$ (indicated by the brown dot-dashed line)~\cite{Volker1998,Weiss1999,Saito2013}.
The agreement between the numerical results and the exact solution indicates the correctness of the CTQMC simulation.

At low temperatures ($k_{\rm B} T \ll \hbar \Delta_{\rm eff}$), the numerical results agree well with those of the approximate formula for the co-tunneling process [equation (\ref{eq:cotunneling2}); indicated by blue dashed lines in figure~\ref{fig:ohmic_conductance}].
In this regime, the thermal conductance is always proportional to $T^3$ ($=T^{2s+1}$), which is consistent with both results of a previous study~\cite{Saito2013}.

At moderate ($k_{\rm B}T \sim \hbar \Delta_{\rm eff}$) and high temperatures ($k_{\rm B} T \gg \hbar \Delta_{\rm eff}$), the numerical results deviate from the co-tunneling formula and agree well with NIBA (indicated by black solid lines in figure~\ref{fig:ohmic_conductance}).
Note that the thermal conductance obtained by NIBA is proportional to $T^{3-2\alpha}$ at low temperatures, as shown in figure~\ref{fig:ohmic_conductance}.
NIBA agrees well even with the low-temperature numerical results for the weak system-reservoir coupling ($\alpha \ll 1$), whereas it deviates from these results as this coupling becomes large. 
It is remarkable that NIBA agrees well with the numerical results at arbitrary temperatures for $\alpha \ll 1$, as shown in figure~\ref{fig:ohmic_conductance}~(a).

In figures~\ref{fig:ohmic_conductance}~(a) and (b), we also show the approximate formula for sequential tunneling (indicated by green dot-dashed lines).
As shown in this figure, the sequential tunneling formula at moderate temperatures ($k_{\rm B} T \sim \hbar \Delta_{\rm eff}$) agrees with the numerical results of the weak system-reservoir coupling ($\alpha \ll 1$).
However, note that NIBA agrees with the numerical results for a wider temperature region than the sequential tunneling formula.

The Markov approximation for incoherent tunneling, indicated by orange dotted lines in figure~\ref{fig:ohmic_conductance}, clearly deviates from the numerical results for $\alpha = 0.05$, $0.1$, and $0.7$, indicating the importance of the non-Markovian properties of the system.
The Toulouse point $\alpha = 0.5$ is an exception, as shown in figure~\ref{fig:ohmic_conductance}~(c); NIBA coincides with the Markov approximation since at this point the self-energy in NIBA becomes independent of the frequency for the unbiased case~\cite{Weiss1999}.
A detailed discussion on the failure of the Markov approximation is given in section~\ref{sec:ResultSubOhmic}.

\begin{figure}[tbp]
	\centering
	\includegraphics[height=5.0cm]{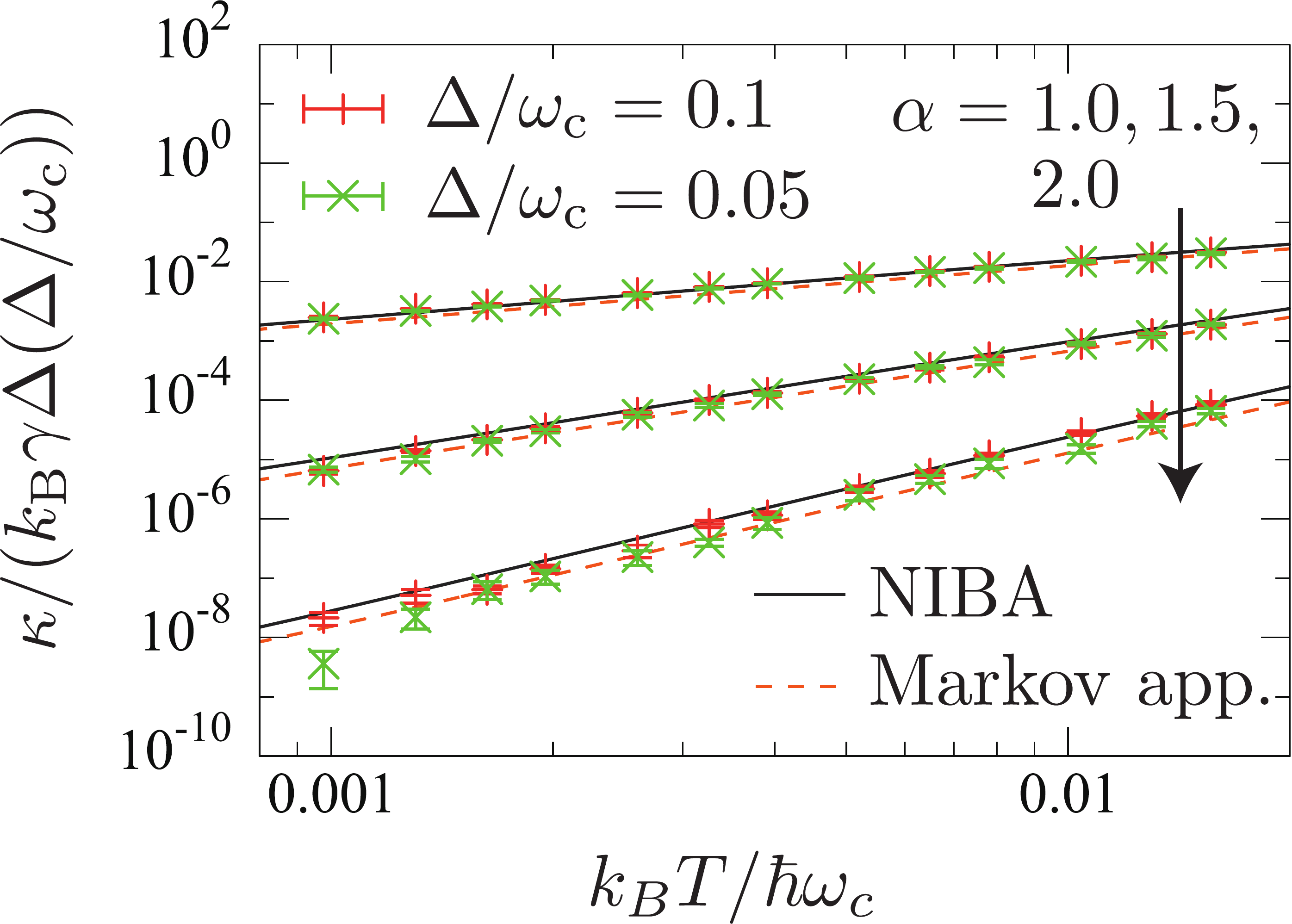}
	\caption{Temperature dependence of the thermal conductance.
    The symbols with error bars indicate the numerical results obtained using the CTQMC method for $\alpha = 1.0$, $1.5$, and $2.0$ from top to bottom. 
 The black solid and orange dashed lines represent NIBA and the Markov approximation for incoherent tunneling.}
	\label{fig:ohmic_incoherent}
\end{figure} 

As described in section~\ref{sec:EffectiveTunnelingAmplitude}, quantum phase transition occurs at $\alpha_{\rm c} = 1$ for the ohmic case.
For $\alpha_{\rm c} \ge 1$, the effective tunneling amplitude $\Delta_{\rm eff}$ becomes zero, indicating complete destruction of the superposition of the two localized states.
Therefore, heat transport is induced by incoherent tunneling at arbitrary temperatures.
In figure~\ref{fig:ohmic_incoherent}, we show the thermal conductance for $\alpha = 1.0$, $1.5$, and $2.0$ as a function of temperature.
As indicated by the black solid lines in the figure, the numerical results agree well with NIBA formula for arbitrary temperatures.
Note that for $\alpha \ge 1$, the condition for the co-tunneling regime $k_{\rm B}T \ll \hbar \Delta_{\rm eff}$ is never satisfied.
In figure ~\ref{fig:ohmic_incoherent}, we also show the Markov approximation for incoherent tunneling (indicated by the orange dashed line).
For $\alpha \ge 1$, the difference between NIBA and the Markov approximation is not considerably large.

\subsection{The sub-ohmic case ($s<1$)}
\label{sec:ResultSubOhmic}

\begin{figure}[tbp]
	\centering
	\includegraphics[height=5.0cm]{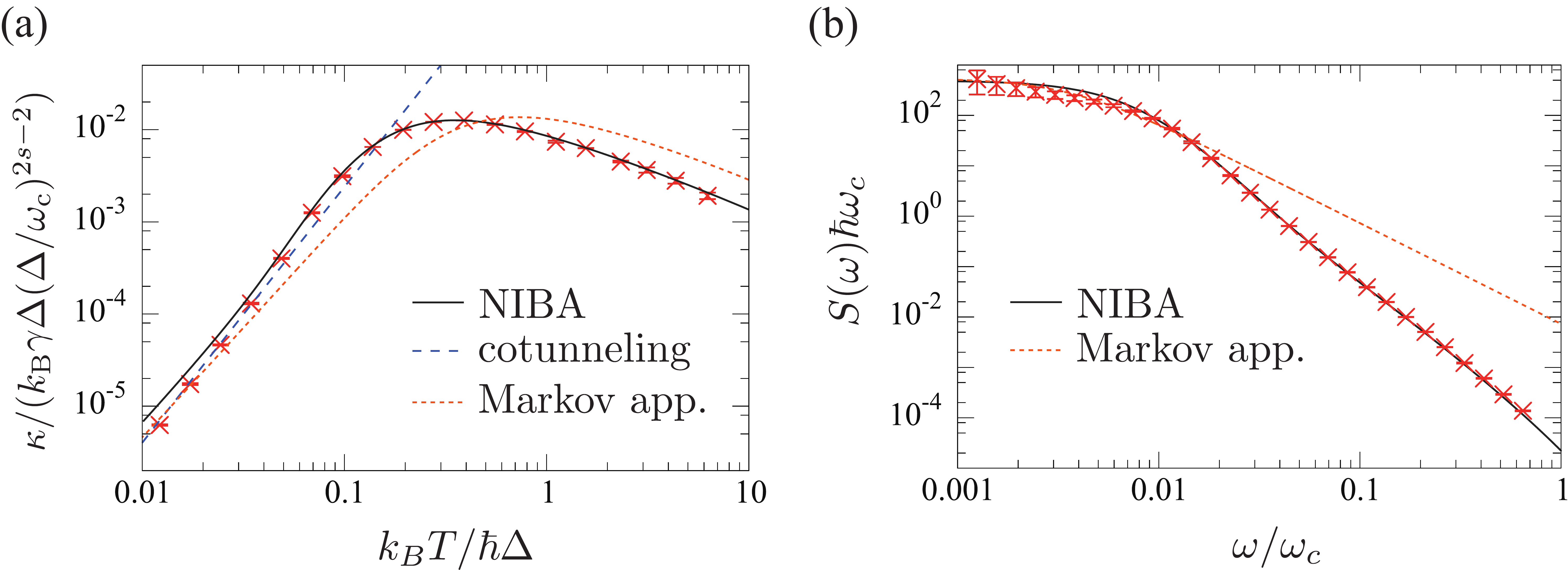}
	\caption{(a) The thermal conductance calculated using the CTQMC method for $s = 0.9$, $\Delta/\omega_c = 0.01$ and $\alpha$=0.1. (b) The symmetrized correlation function calculated using the CTQMC method at $k_{\rm B}T = \hbar\omega_{\rm c}/64$ for parameters same as those considered in (a). 
The black solid, blue dashed, and orange dotted lines represent NIBA, co-tunneling formula, and Markov approximation for incoherent tunneling, respectively.}
	\label{fig:sub_ohmic}
\end{figure}

We first discuss the thermal conductance for the sub-ohmic case wherein the system-reservoir coupling is below the critical value for the quantum phase transition.
In figure~\ref{fig:sub_ohmic}~(a), we show the thermal conductance as a function of the temperature for $s=0.9$, $\Delta/\omega_c=0.01$, and $\alpha = 0.1$, for which the ground state is delocalized ($\alpha < \alpha_c(s,\Delta)$).
At moderate and high temperatures, the numerical results agree well with the NIBA, which is shown by the black solid line.
We note that the sequential-tunneling formula cannot be applied to the sub-ohmic case.
At low temperatures ($k_{\rm B} T \ll \hbar \Delta_{\rm eff}$), the numerical results agree well with the co-tunneling formula, showing $T^{2s+1}$-dependence.

We also show the results of the Markov approximation for incoherent tunneling by the orange dotted line in figure~\ref{fig:sub_ohmic}~(a).
The Markov approximation clearly deviates from the numerical results. 
To understand the failure of the Markov approximation,
we show the numerical and analytical result of the symmetrized correlation function $S(\omega)$ as a function of $\omega/\omega_c$ for $k_{\rm B}T = \hbar\omega_{\rm c}/64$ in figure~\ref{fig:sub_ohmic}~(b).
While the Markov approximation for the incoherent tunneling process agrees with the numerical results at a low frequency, clear deviation is observed at higher frequencies;
the numerical result indicates that the high-frequency decay of $S(\omega)$ is much faster than that of the Markov approximation, which is proportional to $\omega^{-2}$ (see equation (\ref{eq:SomegaIncoherent}))
We note that the numerical result of $S(\omega)$ is well reproduced by the NIBA at arbitrary frequencies.
These observations indicate that the non-Markovian properties of the system dynamics are important for obtaining correct thermal conductance results for the sub-ohmic case.

\begin{figure}[tbp]
	\centering
	\includegraphics[height=5.0cm]{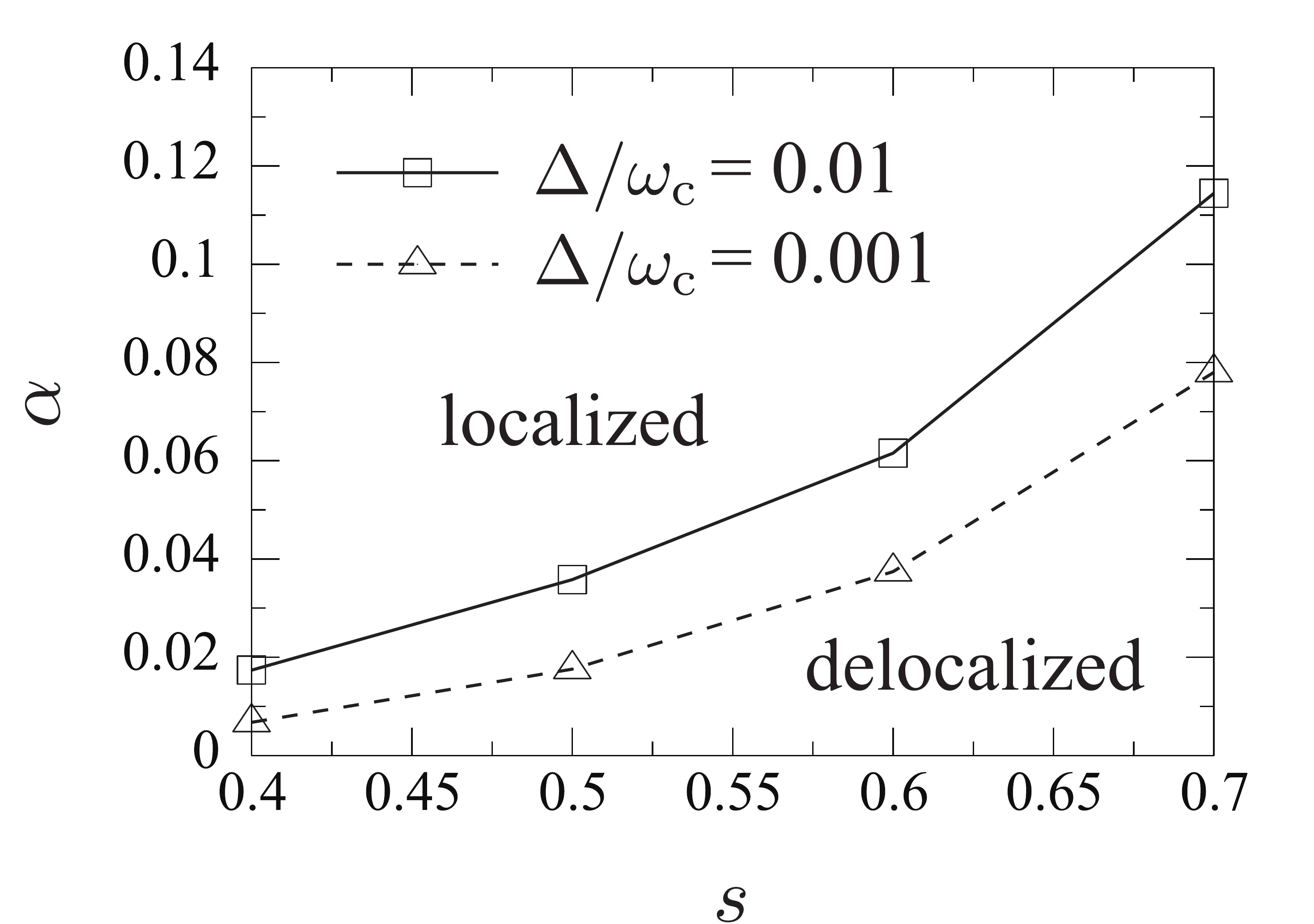}
	\caption{Phase diagram for the transition between the delocalized and localized phases.
    The square and triangle symbols indicate the critical values of the system-reservoir coupling, 
    $\alpha_c$ for $\Delta/\omega_c=0.01$ and $0.001$, respectively.}
	\label{fig:phase_diagram}
\end{figure}

Next, let us study the effect of the quantum phase transition.
Figure \ref{fig:phase_diagram} shows the phase diagram determined by the CTQMC method.
The detailed procedure for the determination of the critical point is given in \ref{app:Binder}.
The obtained critical system-reservoir coupling, $\alpha_c$, for the quantum phase transition is a function of both $s$ and $\Delta$ and is consistent with previous work based on the NRG calculation~\cite{Bulla2003}.

\begin{figure}[tbp]
	\centering
	\includegraphics[height=5.0cm]{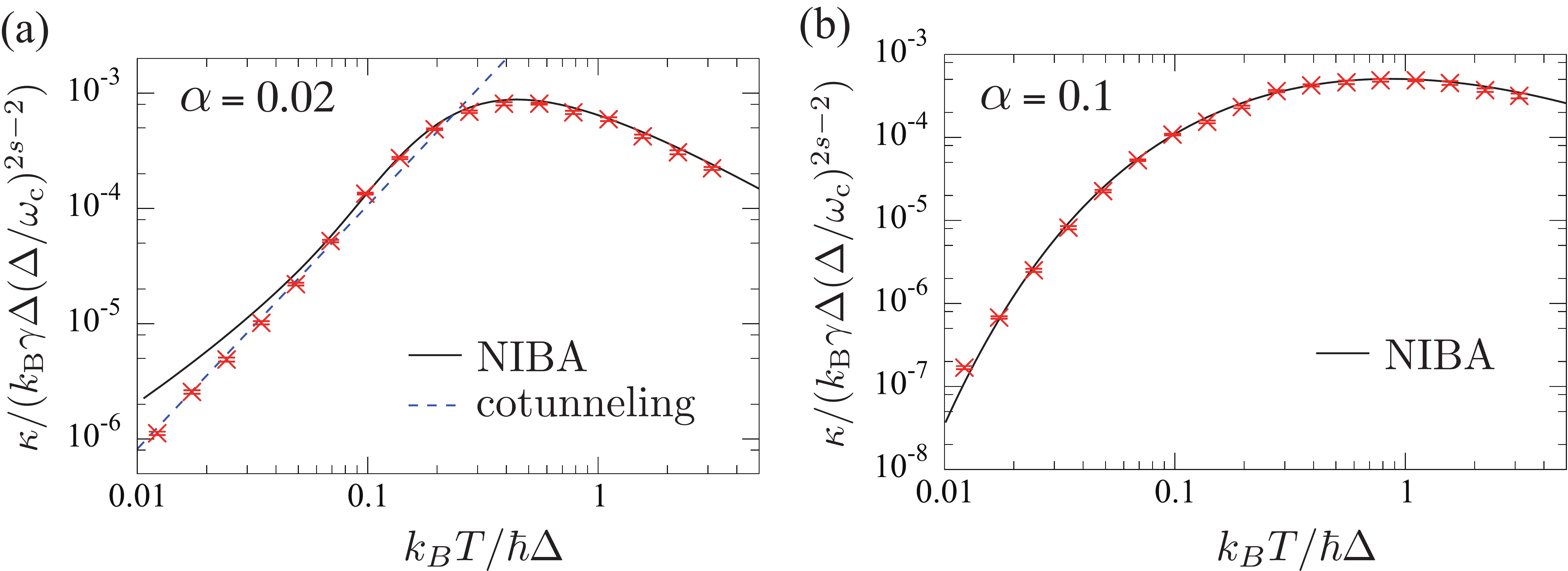}
	\caption{Temperature behavior of the thermal conductance calculated by a Monte Carlo simulation. The data represent results for $s = 0.6$, $\Delta/\omega_{\rm c} = 0.01$, (a) $\alpha = 0.02$, and (b) $\alpha = 0.1$.  
  The black solid lines and the blue dashed line represents NIBA and the co-tunneling formula, respectively.}
	\label{fig:sub_ohmic_QPT}
\end{figure}

The quantum phase transition remarkably affects the temperature dependence of the thermal conductance.
In figure~\ref{fig:sub_ohmic_QPT}, we show the thermal conductance as a function of the temperature for $s=0.6$ and $\Delta/\omega_{\rm c} = 0.01$, for which a quantum phase transition occurs at $\alpha = \alpha_{\rm c} = 0.0615$.
Figure~\ref{fig:sub_ohmic_QPT}~(a) shows the temperature dependence in the delocalized regime ($\alpha = 0.02 < \alpha_{\rm c}$), for which $\Delta_{\rm eff}$ remains finite.
The numerical results agree well with the co-tunneling formula at low temperatures and with NIBA at moderate-to-high temperatures.
This feature is the same as that shown in figure~\ref{fig:sub_ohmic}.
Figure~\ref{fig:sub_ohmic_QPT}~(b) shows the temperature dependence in the localized regime ($\alpha = 0.1 > \alpha_{\rm c}$), for which $\Delta_{\rm eff}=0$.
Reflecting the quantum phase transition, the numerical results agree with NIBA at arbitrary temperatures, as shown in figure~\ref{fig:sub_ohmic_QPT}~(b).
Since the condition for the co-tunneling regime, $k_{\rm B} T \ll \hbar \Delta_{\rm eff}$, is never satisfied for $\Delta_{\rm eff}=0$, the thermal conductance does not show a universal $T^{2s+1}$-dependence due to the co-tunneling process at low temperatures.

\subsection{The super-ohmic case ($1<s<2$)}
\label{sec:ResultSuperOhmic}

\begin{figure}[tbp]
	\centering
	\includegraphics[height=5.0cm]{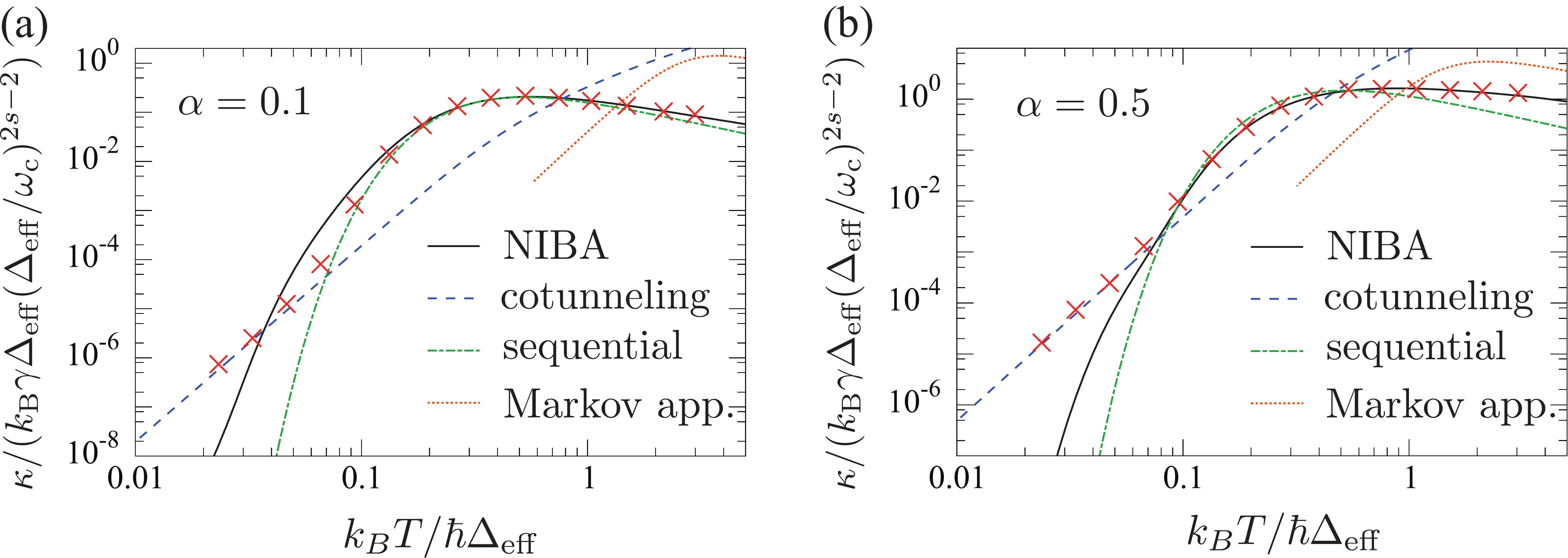}
	\caption{Temperature behavior of the thermal conductance calculated by Monte Carlo simulation. The data represent results for $s$ = 1.5, $\Delta/\omega_{\rm c} = 0.05$, (a) $\alpha = 0.1$, and (b) $\alpha = 0.5$. 
    In both figures, the black solid, blue dashed, green dot-dashed, and orange dotted lines represent NIBA, co-tunneling formula, sequential tunneling formula, and Markov approximation for incoherent tunneling, respectively.}
	\label{fig:super_ohmic_s1.5}
\end{figure} 

In figure~\ref{fig:super_ohmic_s1.5}, we show the numerical thermal conductance results obtained using CTQMC as a function of temperature for $s=1.5$.
Here, the horizontal and vertical axes are the normalized temperature $k_{\rm B}T/\hbar\Delta_{\rm eff}$ and the normalized thermal conductance $\kappa/(k_{\rm B} \gamma \Delta_{\rm eff} (\Delta_{\rm eff}/\omega_c)^{2s-2})$, respectively, where $\Delta_{\rm eff}$ is the effective tunneling amplitude defined in equation~(\ref{delta_eff_super_ohmic}).
Note that there is no quantum transition for the super-ohmic case ($s>1$); $\Delta_{\rm eff}$ is finite for arbitrary system-reservoir couplings.
At low temperatures ($k_{\rm B}T \ll \hbar \Delta_{\rm eff}$), the numerical results agree with the co-tunneling formula (indicated by blue dashed lines) and show $T^{2s+1}$-dependence, regardless of the strength of the system-reservoir coupling.
As shown in figure~\ref{fig:super_ohmic_s1.5}~(a), the numerical results for $\alpha = 0.1$ agree with the sequential tunneling formula at moderate temperatures ($k_{\rm B}T \sim \hbar \Delta_{\rm eff}$) and with NIBA at high temperatures.
However, from figure~\ref{fig:super_ohmic_s1.5}~(b), it is evident that the numerical results for $\alpha=0.5$ agree better with NIBA than with the sequential tunneling formula at moderate-to-high temperatures ($k_{\rm B}T \gtrsim \hbar \Delta_{\rm eff})$.
This change can be explained by the crossover temperature $T^*$, which separates the sequential ($T<T^*$) and incoherent ($T>T^*$) tunneling regimes [see equation~(\ref{eq:CrossoverTemperature})].
As the system-reservoir coupling $\alpha$ increases, the temperature region for which the numerical results agree with NIBA is widened since the crossover temperature $T^*$ is lowered.

The Markov approximation for incoherent tunneling is indicated by orange dotted lines in figure~\ref{fig:super_ohmic_s1.5}. 
The incoherent tunneling formula clearly deviates from numerical results, indicating the importance of the non-Markovian properties of the system dynamics.
The origin of this disagreement is same as that for the sub-ohmic case (see section~\ref{sec:ResultSubOhmic}).

\subsection{The super-ohmic case ($s \ge 2$)}
\label{sec:ResultSuperOhmic2}

\begin{figure}[tbp]
	\centering
	\includegraphics[height=5.0cm]{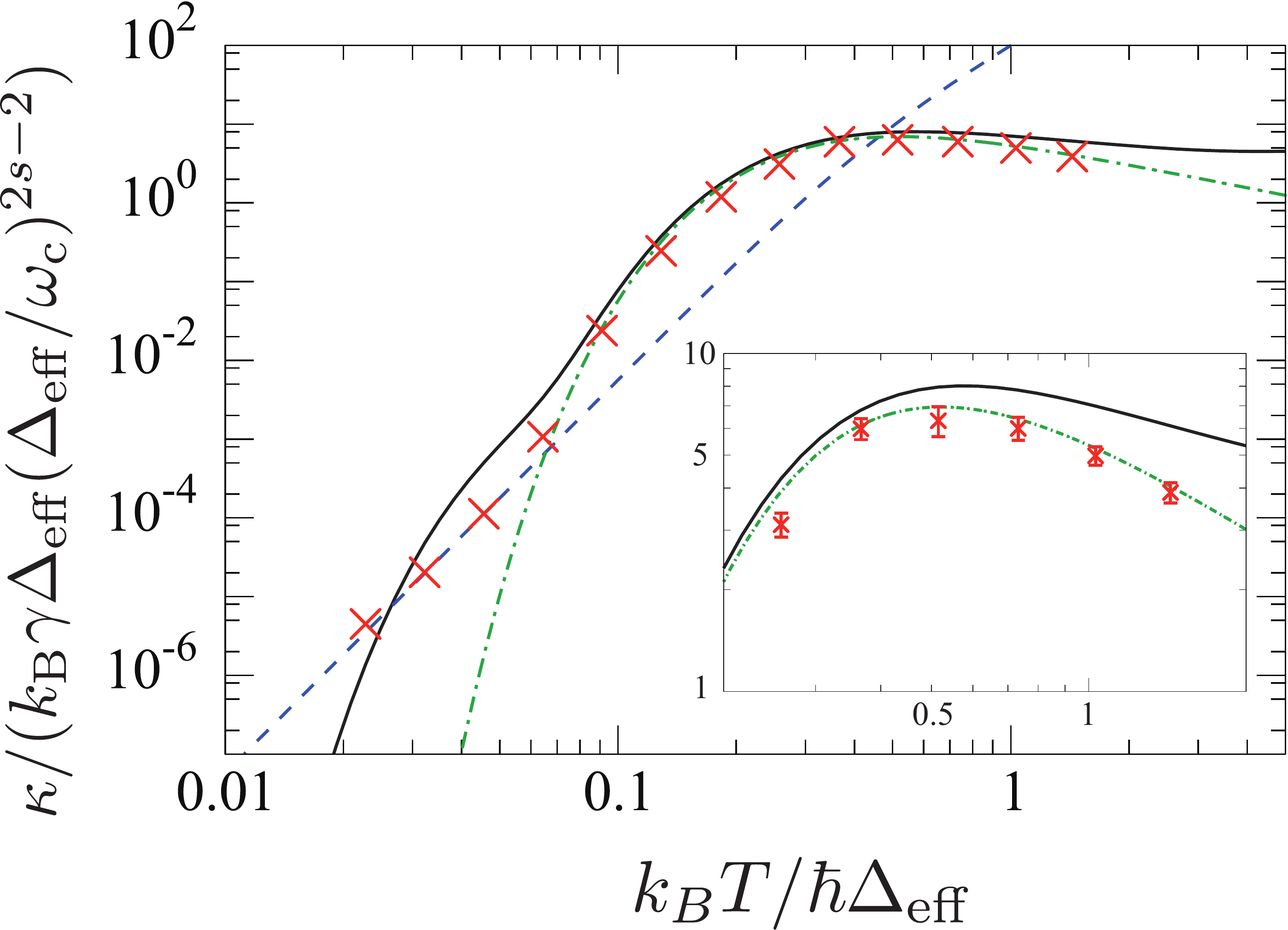}
	\caption{Temperature dependence of the thermal conductance calculated using CTQMC simulation. The data represent the result for $s = 2.0$, $\Delta/\omega_{\rm c} = 0.05$, and $\alpha$ = 0.5.
The linetypes are same as those in figure~\ref{fig:super_ohmic_s1.5}.
The inset shows an enlarged graph in the high-temperature region.}
	\label{fig:super_ohmic_s2.0}
\end{figure} 

In figure~\ref{fig:super_ohmic_s2.0}, we show the numerical results of the thermal conductance obtained using the CTQMC method as a function of the temperature for $s=2.0$.
The normalization of the horizontal and vertical axes as well as the linetypes of the analytical formula are same as those in figure~\ref{fig:super_ohmic_s1.5}.
At low temperatures, the numerical results agree well with the co-tunneling formula and show $T^{2s+1}$-dependence, regardless of the strength of the system-reservoir coupling.
In contrast to the case of $1<s<2$, the numerical results agree with the sequential tunneling formula at moderate-to-high temperatures.
This is reasonable since the crossover $T^*$ becomes of the order of $\omega_c$ for $s=2$.

\section{Summary}

We systematically considered heat transport via a local two-state system for all types of reservoirs, i.e., for the ohmic case ($s=1$), super-ohmic case ($s>1$), and sub-ohmic case ($s<1$). We used the exact expression for the thermal conductance obtained from the Keldysh formalism and studied it using both analytic and numerical methods.

First, we considered the approximations of three transport processes: sequential tunneling, co-tunneling, and incoherent tunneling.
In particular, we newly derived a universal formula for co-tunneling using the generalized Shiba relation, which predicts the $T^{2s+1}$-dependence of the thermal conductance at low temperatures.
We also pointed out that the Markov approximation yielded incorrect results for the thermal conductance in the incoherent tunneling regime since the non-Markovian properties are important.
However, for the incoherent tunneling regime, NIBA yielded correct results.

Next, we used a continuous-time Monte Carlo algorithm and systematically compared the numerical results with those of the analytical approximation formulas.
We found that all numerical results were well reproduced by one of three formulas, i.e., the sequential tunneling formula, co-tunneling formula, or NIBA.
The formulas that yielded correct results are summarized in Table~\ref{table:tunneling_process}.
We also showed that for $0<s\le 1$, the quantum phase transition between the delocalized and localized phases strongly affected the temperature dependence of the thermal conductance.
For the delocalized phase ($\alpha < \alpha_{\rm c}$), the thermal conductance is well described by the co-tunneling formula at low temperatures and by NIBA at moderate-to-high temperatures.
On the contrary, for the localized phase ($\alpha > \alpha_{\rm c}$), NIBA holds at arbitrary temperatures.

Our study is expected to provide a theoretical basis for describing heat transport via nano-scale objects.
Herein, we focused on heat transport in a symmetric double-well-shaped potential near the thermal equilibrium in the limit of $\Delta \ll \omega_{\rm c}$.
The effect of asymmetry of system's potential, the cutoff-frequency dependence, and the far-form-equilibrium effect constitute
an important future problem.
The temperature dependence of the thermal conductance in the critical regime near the quantum phase transition is also an intriguing subject for research and will be discussed elsewhere.

\section*{Acknowledgement}
The authors thank R. Sakano and T. Yokoyama for helpful discussions and comments.
T.K. was supported by JSPS Grants-in-Aid for Scientific Research (No. JP24540316 and JP26220711).
K.S. was supported by JSPS Grants-in-Aid for Scientific Research (No. JP25103003, JP16H02211, and JP17K05587).
\appendix
\section{Derivation of the Meir-Wingreen-Landauer Formula}
\label{app:MierWingreeen}

In this appendix, based on previous research~\cite{Ojanen2008,Ruokola2009,Saito2008,Saito2013}, we derive the Meir-Wingreen-Landauer-type formula~\cite{Meir1992} given by equation (\ref{eq:exact_current}) for the heat current in the Keldysh formalism.
We define the nonequilibrium Green function as~\cite{Rammer1984,Jauho1994,Jauho2007}
\begin{equation}
G_{A,B}(u,u^\prime) = -\frac{i}{\hbar} \langle T_K A(u) B(u^\prime) \rangle,
\end{equation}
where $A$ and $B$ are bosonic operators, $u$ is a time variable on the Keldysh contour comprising the forward and backward paths, $T_K$ is a time-ordered product on the Keldysh contour. 
The average indicated by $\langle \cdots \rangle$ is taken for the initial-state density matrix
\begin{eqnarray}
& & \rho = \rho_S \prod_{\nu={\rm L},{\rm R}} \otimes \rho_\nu, \label{eq:initial1} \\
& & \rho_\nu = e^{-\beta_\nu \sum_k \hbar\omega_{\nu k} b_{\nu k}^\dagger b_{\nu k}}/Z_\nu,
\label{eq:initial2}
\end{eqnarray}
at $t= -\infty$, where $Z_\nu$ is the partition function of the isolated reservoir $\nu$.
By projection from the Keldysh contour onto the real-time axis, the retarded, advanced, and lesser components of the nonequilibrium Green function are respectively defined as
\begin{eqnarray}
& & G^{r}_{A,B}(t,t^\prime) = - \frac{i}{\hbar} \theta(t-t^\prime) \langle [B(t^\prime),A(t)]\rangle, \\
& & G^{a}_{A,B}(t,t^\prime) = \frac{i}{\hbar} \theta(t^\prime-t)\langle [B(t^\prime),A(t)]\rangle, \\
& & G^{<}_{A,B}(t,t^\prime) = -\frac{i}{\hbar} \langle B(t^\prime)A(t)\rangle, 
\end{eqnarray}
where $\theta(t)$ is the Heaviside step function.

The nonequilibrium steady-state heat current is written in terms of the Keldysh Green function as 
\begin{eqnarray}
\langle J_\nu(t) \rangle = \sum_k {\rm Re}\left[ \hbar^2 \lambda_{\nu k}
 \omega_{\nu k} G^{<}_{\sigma_z,b^{\dagger}_{\nu,k}}(t,t) \right].
\end{eqnarray}
For the initial state given in equations~(\ref{eq:initial1}) and (\ref{eq:initial2}), one can derive the relation 
\begin{equation}
G_{\sigma_z,b^\dagger_{\nu k}}(u,u^\prime) = \frac{\hbar \lambda_{\nu k}}{2}
\int du_1 G_{\sigma_z,\sigma_z}(u,u_1) g_{b_{\nu k},b^{\dagger}_{\nu k}}(u_1,u^\prime),
\label{eq:keldyshidentity}
\end{equation}
using the formal expansion with respect to $\lambda_{\nu k}$, where $g_{b_{\nu k},b^{\dagger}_{\nu k}}(u,u^\prime)$ is the Green function for the isolated reservoir $\nu$ and integration with respect to $u_1$ is performed on the Keldysh contour.
By projection onto the real-time axis, the lesser component of equation~(\ref{eq:keldyshidentity}) is rewritten as
\begin{eqnarray}
& & G^<_{\sigma_z,b^\dagger_{\nu k}}(t,t^\prime) 
= \frac{\hbar \lambda_{\nu k}}{2}
\int_{-\infty}^{\infty} dt_1 
\biggl[ G^r_{\sigma_z,\sigma_z}(t,t_1) g^<_{b_{\nu k},b^{\dagger}_{\nu k}}(t_1,t^\prime) 
\nonumber \\
& & \hspace{45mm} 
+ G^<_{\sigma_z,\sigma_z}(t,t_1) g^a_{b_{\nu k},b^{\dagger}_{\nu k}}(t_1,t^\prime) \biggr].
\label{eq:keldyshidentity2}
\end{eqnarray}
The heat current is then rewritten as
\begin{eqnarray}
& & \langle J_\nu(t) \rangle = \lim_{t^\prime \rightarrow t} 
2 \, {\rm Re} \int_{-\infty}^\infty dt_1(-i\hbar \partial_{t^\prime})
\biggl[ G^r_{\sigma_z,\sigma_z}(t,t_1) \Sigma_\nu^<(t_1,t^\prime) \nonumber \\
& & \hspace{55mm} +  G^<_{\sigma_z,\sigma_z}(t,t_1) \Sigma_\nu^a(t_1,t^\prime) 
\biggr], 
\label{eq:currentexpression1}
\end{eqnarray}
where $\Sigma^<_\nu(t,t^\prime)$ and $\Sigma^a_\nu(t,t^\prime)$ are the lesser and advanced components, respectively, of the reservoir self-energy
\begin{equation}
\Sigma_\nu(u,u^\prime) = \sum_{k} \frac{(\hbar \lambda_{\nu k})^2}{4} g_{b_{\nu k},b^{\dagger}_{\nu k}}(u,u^\prime),
\end{equation}
which are calculated as
\begin{eqnarray}
& & \Sigma^<_\nu(t,t^\prime) = - \frac{i}{4} \int_0^\infty d(\hbar \omega) \, I_\nu(\omega) n_\nu(\omega) e^{-i\omega(t-t^\prime)}, \\
& & \Sigma^a_\nu(t,t^\prime) = \frac{i}{4} \theta(t^\prime - t) \int_0^\infty d(\hbar \omega) \, I_\nu(\omega) e^{-i\omega(t-t^\prime)},
\end{eqnarray}
respectively.
Here, $n_\nu(\omega) = (e^{\hbar \omega/k_{\rm B} T_{\nu}}-1)^{-1} $ is the Bose distribution function of phonons (photons) for reservoir $\nu$.
The Fourier transformation of equation~(\ref{eq:currentexpression1}) gives
\begin{eqnarray}
	\average{J_\nu}=\frac{1}{2} \int_{0}^{\infty}d(\hbar \omega) \, 
		\hbar \omega I_\nu(\omega) \left[ 
			\mathrm{Im}[G_{\sigma_z,\sigma_z}^{r}(\omega)]n_\nu(\omega)-
			\frac{i}{2} G^{<}_{\sigma_z,\sigma_z}(\omega)
		\right],
\end{eqnarray}
where $G_{\sigma_z,\sigma_z}^{r}(\omega)$ and $G^{<}_{\sigma_z,\sigma_z}(\omega)$ are the Fourier transformations of the retarded and lesser components of the nonequilibrium Green function, respectively.
Considering the conservation law of energy given by $\average{J_L} = -\average{J_R}\equiv \average{J}$, the heat current is rewritten as
\begin{eqnarray}
\average{J} &=& \frac{\alpha_R}{\alpha_L + \alpha_R} \langle J_L \rangle - \frac{\alpha_L}{\alpha_L + \alpha_R} \langle J_R \rangle \nonumber \\
&=& \frac{\alpha_L \alpha_R}{2(\alpha_L+\alpha_R)} \int_{0}^{\infty}d(\hbar \omega) \, 
		\hbar \omega\,
			\mathrm{Im}[G_{\sigma_z,\sigma_z}^{r}(\omega)]\tilde{I}_\nu(\omega) [n_L(\omega)-n_R(\omega)].
\end{eqnarray}
Here, we used $I_\nu(\omega) = \alpha_\nu \tilde{I}(\omega)$.
Rewriting $G_{\sigma_z,\sigma_z}^{r}(\omega)$ with $\chi(\omega)$, we finally obtain equation (\ref{eq:exact_current}).

\section{Adiabatic Renormalization}
\label{app:AdiabaticRenormalization}

We consider oscillators in the reservoirs whose frequencies are in the range $p\omega_c < \omega < \omega_c$, where the factor $p$ is first simply assumed to be slightly smaller than 1.
For the zeroth-order adiabatic approximation, we assume that these high-frequency oscillators ($\Delta \ll p\omega_c \sim \omega_c$) instantaneously adjust their quantum states to the current value of $\sigma_z$.
If, for a moment, we ignore the other low-frequency oscillators, the wavefunctions of the two lowest energy eigenstates for the system-plus-reservoir are described by
\begin{eqnarray}
& & \Ket{E_0^\prime}= \frac{1}{\sqrt{2}} \left( \Ket{\Psi_{\rm L}}+ \Ket{\Psi_{\rm R}} \right), \\
& & \Ket{E_1^\prime} = \frac{1}{\sqrt{2}} \left( \Ket{\Psi_{\rm L}}- \Ket{\Psi_{\rm R}} \right), 
\end{eqnarray}
where $\Ket{\Psi_{\rm L}}$ and $\Ket{\Psi_{\rm R}}$ are given by
\begin{eqnarray}
& &	\Ket{\Psi_{\rm L}} =\Ket{\sigma_z = -1}
    \otimes\prod_{\nu k}{\vphantom\prod}' \Ket{\Psi_{\nu k}^-}, \\
& & \Ket{\Psi_{\rm R}} =\Ket{\sigma_z = +1} 
    \otimes\prod_{\nu k}{\vphantom\prod}' \Ket{\Psi_{\nu k}^+},
\end{eqnarray} 
respectively.
Here, the prime symbol indicates that the product is in the range $p\omega_c < \omega_{\nu k} < \omega_c$. 
$\Ket{\Psi_{\nu k}^\pm}$ is the ground-state wave function of the oscillator $k$ in reservoir $\nu$ when the wavefunction of the local system is located at $x = \pm x_0/2$; it is obtained by translation of the ground-state wavefunction $\Ket{\Psi_{\nu k}^0}$ for the isolated oscillator as
\begin{eqnarray}
& & \Ket{\Psi_{\nu k}^\pm} = 
\exp \left( \pm \frac{i}{\hbar} \delta_{\nu k} p_{\nu k} \right)
 \Ket{\Psi_{\nu k}^0}, \\
 & & \delta_{\nu k} = -\frac{C_{\nu k}}{m_{\nu k}\omega_{\nu k}^2} \frac{x_0}{2}.
\end{eqnarray}
Adiabatic renormalization suggests that the tunneling amplitude is renormalized by the overlap between the ground-state wavefunctions of the oscillators for different localized states ($\sigma_z =\pm1$):
\begin{eqnarray}
\Delta^\prime(p) =\Delta\prod_{\nu k}{\vphantom\prod}'  \Braket{\Psi_{\nu k}^+| \Psi_{\nu k}^-}
\simeq \Delta \exp\left(-\alpha\int_{p\omega_c}^{\omega_c} d\omega\ 
\frac{(\omega/\omega_c)^{s-1}}{\omega}\right).
\label{eq:AdiabaticRenormalization}
\end{eqnarray}
If the renormalized tunneling amplitude $\Delta'(p)$ is less than $p\omega_c$, the adiabatic renormalization can continue by reducing the factor $p$.
If $\Delta'(p^*)=p^*\omega_c$ holds at $p=p^*$, adiabatic renormalization must be stopped there and the finite effective tunneling amplitude $\Delta_{\rm eff} = \Delta^\prime(p^*)$ is obtained.
On the contrary, if $\Delta'(p) < p \omega_c$ holds for an arbitrary value of $p$, adiabatic renormalization can be completed even at $p=0$, yielding an effective tunneling amplitude of zero ($\Delta_{\rm eff} =0$).

For the ohmic case ($s=1$), the effective tunneling amplitude is obtained as follows:
\begin{eqnarray}
\Delta_{\mathrm{eff}}^\prime =\left\{ \begin{array}{ll}
\displaystyle{\Delta \left(\frac{\Delta}{\omega_c}\right)^{\alpha/(1-\alpha)}}, & (0\le \alpha<1), \\
0, & (1 \le \alpha ).
\end{array} \right. 
\end{eqnarray}
In this paper, following Ref.~\cite{Weiss1999}, we employ a modified effective tunneling amplitude multiplied by a dimensionless function of $\alpha$:
\begin{eqnarray}
\Delta_\mathrm{eff} \equiv
[\Gamma(1-2\alpha)\cos(\pi\alpha)]^{1/2(1-\alpha)}
\Delta_{\mathrm{eff}}^\prime.
\end{eqnarray}
Using this definition, equation~(\ref{eq:OhmicDeltaEff}) is derived.

Based on equation (\ref{eq:AdiabaticRenormalization}), it is straightforward to show that the effective tunneling amplitude in the super-ohmic case ($s>1$) assumes a finite value given by (\ref{delta_eff_super_ohmic}) and that it always vanishes for the sub-ohmic case ($s<1$).

\section{Continuous-time Quantum Monte Carlo Method}
\label{app:CTQMC}

In early numerical studies~\cite{Chakravarty1995,Volker1998,Umeki2007}, the Monte Carlo method has been applied directly to the long-range Ising model, which is mapped from the spin-boson model~\cite{Weiss1999,Anderson1971,Cardy1981,Luijten1995,Leggett1987}. Subsequently, the continuous-time quantum Monte Carlo (CTQMC) algorithm~\cite{Rieger1999,Gubernatis2016} has been applied directly to the spin-boson model without mapping~\cite{Winter2009}.
In this section, we describe the CTQMC algorithm employed in the present numerical simulation.

\begin{figure}[tbp]
	\centering
	\includegraphics[height=10.0cm]{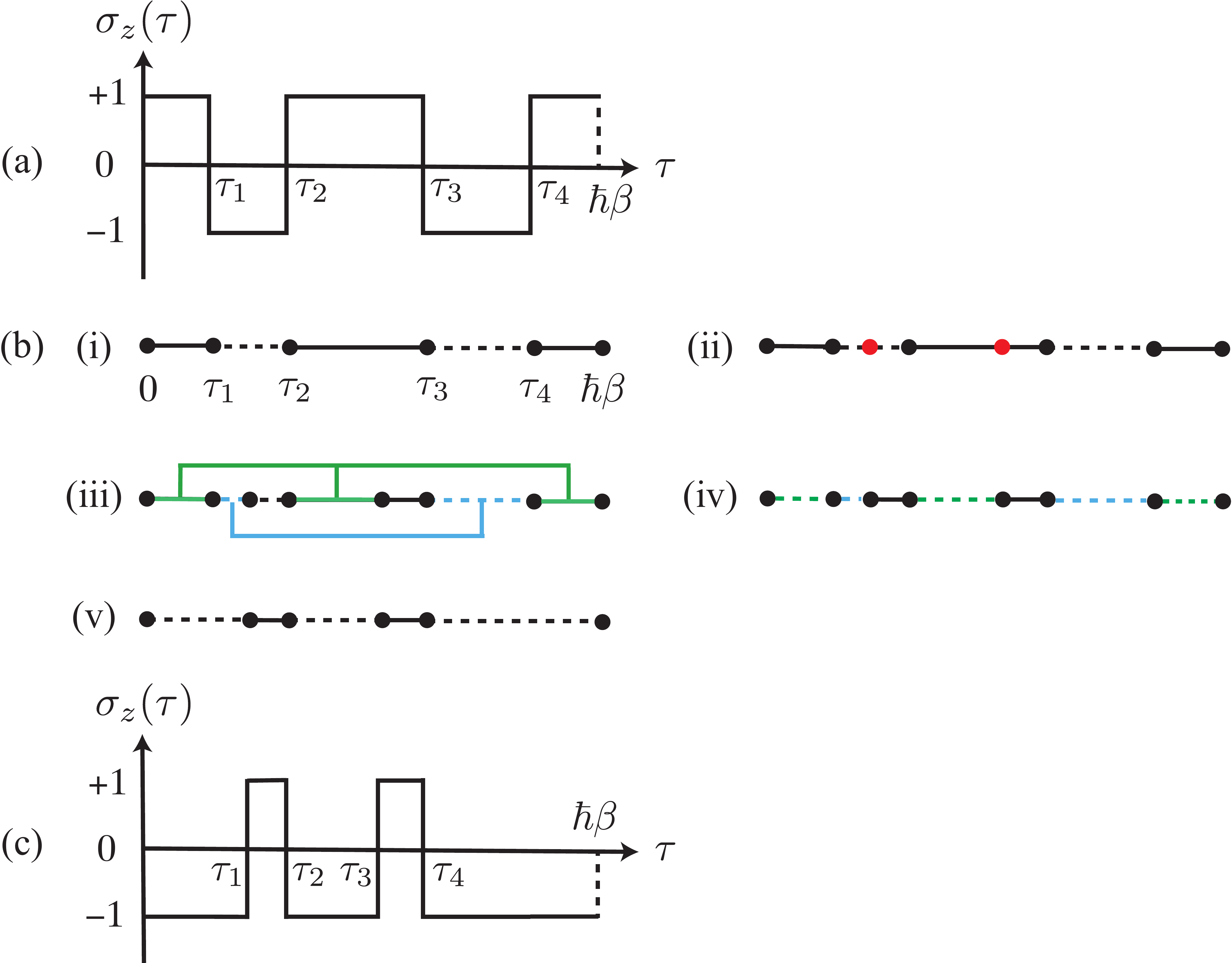}
	\caption{
	 The spin path on the imaginary $\tau$-axis for the spin-boson model
     and the CTQMC update process: (a) An initial spin path,
    (b-i) the vertex representation of (a),
	 (b-ii) insertion of new cuts using the Poisson distribution,
	 (b-iii) connection of segments with the probability (\ref{prob}),
	 (b-iv) flipping each cluster with probability 1/2, 
    (b-v) removal of redundant cuts, and 
    (c) the final spin path after the update.}
	\label{fig:config}
\end{figure}

The partition function of the spin-boson model (\ref{Hamiltonian}) is written in the path-integral form as~\cite{Weiss1999,Winter2009}
\begin{eqnarray}
Z = \int\mathcal{D}\sigma_z (\tau)\ \exp\left[-\frac{1}{4}\int_0^{\hbar \beta} d\tau\int_0^\tau d\tau^\prime \ \sigma_z(\tau)K(\tau-\tau^\prime)\sigma_z(\tau^\prime)\right], 
\label{eq:PartitionFunction}
\end{eqnarray}
where $\sigma_z(\tau) (=\pm 1)$ is a spin variable defined on the imaginary-time axis, 
$\mathcal{D}\sigma_z(\tau)$ indicates the integral for all possible paths $\sigma_z(\tau)$,
and $K(\tau)$ is a kernel defined as
\begin{eqnarray}
K(\tau) = \int_0^\infty d\omega \, I(\omega)\frac{\cosh[\omega(\hbar \beta/2-\tau)]}{\sinh(\hbar\beta\omega/2)}.
\label{eq:Kernel}
\end{eqnarray}
As shown in figure~\ref{fig:config}~(a), the path $\sigma_z(\tau)$ is assigned by an alternative configuration of kinks (jumps from $\sigma_z = -1$ to $\sigma_z = +1$) and anti-kinks (jumps from $\sigma_z = +1$ to $\sigma_z = -1$) and described by the positions $\tau_i$ ($i=1,2,\cdots,2n$) of the kinks ($q_i=+1$) and anti-kinks ($q_i=-1)$ as
\begin{eqnarray}
\frac{d \sigma_z}{d\tau} (\tau) = \sum_{i=1}^{2n} 2 q_i \delta(\tau-\tau_i),
\label{eq:path}
\end{eqnarray}
where $n$ is the number of the pairs of kinks and anti-kinks.
Note that the kinks and anti-kinks are alternatively located ($q_{i+1} = - q_i$).
By substituting equation~(\ref{eq:path}) into equation~(\ref{eq:PartitionFunction}), we obtain 
\begin{eqnarray}
Z = \sum_{n=0}^{\infty} \left( \frac{\Delta}{2} \right)^{2n} \int_0^{\hbar\beta} d\tau_{2n} \cdots \int_0^{\tau_2} d\tau_1 \,
\exp\left[\sum_{\langle i , j\rangle}^{2n} q_i q_j W(\tau_i-\tau_j) \right] , 
\label{eq:PartitionFunction2}
\end{eqnarray}
where $\Delta$ is a tunneling matrix element and $W(\tau)$ is obtained from the relation $W''(\tau) = -K(\tau)$ as
\begin{eqnarray}
W(\tau) = \int_0^\infty d\omega \, \frac{I(\omega)}{\omega^2} \frac{\cosh(\hbar\beta\omega/2)-\cosh[\omega(\hbar\beta/2-\tau)]}{\sinh(\hbar\beta\omega/2)}.
\end{eqnarray}
Here, we apply the CTQMC method to this partition function.
The present CTQMC algorithm~\cite{Rieger1999} employs a cluster-flip update similar to that in the Swendsen-Wang cluster algorithm~\cite{Swendsen1987}.
The cluster-flip update is constructed as follows~\cite{Winter2009} (see figure~\ref{fig:config}).
We consider the initial path $\sigma_z(\tau)$ of figure~\ref{fig:config}~(a), and express it via segment representation, as in~(b-i). 
We first insert new vertices with Poisson statics given by $P(\Delta\tau)=\Gamma\exp(-\Gamma\Delta\tau)$ with the mean value $\Gamma^{-1}=2/\Delta$, as shown in (b-ii).
Next, we define the segments $s_i$ (the line segments between neighboring vertices) and connect two segments, $s_i$ and $s_j$, with the probability
\begin{eqnarray}
& & p[s_i,s_j]=1-\delta_{\sigma_z(s_i),\sigma_z(s_j)}[1-e^{-2A}], \\
& & A=W(\tau_{i-1}-\tau_{j-1})-W(\tau_{i-1}-\tau_{j})-W(\tau_{i}-\tau_{j-1})+W(\tau_{i}-\tau_{j}), 
\label{prob}
\end{eqnarray} 
as shown in~(b-iii), and construct segment clusters.
Here, $\sigma_z(s_i)$ is the value of $\sigma_z$ in the segment $s_i$ and the positions of the vertices (including the inserted ones) at the two edges of the segment $s_i$ are denoted by $\tau_{i-1}$ and $\tau_i$, respectively.
Finally, we flip each segment cluster with probability $1/2$, as shown in~(b-iv), and remove the redundant vertices within segments, as shown in (b-v).
The final path is then given by figure~\ref{fig:config}~(c).
The Monte Carlo data presented in this paper typically represent averages over $10^3$-$10^4$ updates at low temperatures and $10^7$-$10^8$ updates at high temperatures. 

Using the CTQMC method, we evaluate the spin correlation function $C(i\omega_n)$ defined in equation~(\ref{eq:CMatsubara}) using the Monte Carlo sampling method as follows:
\begin{eqnarray}
C(i\omega_n) = \frac{1}{\hbar\beta\omega_n^2}\Braket{|\rho(i\omega_n)|^2}, 
\end{eqnarray}
where $\Braket{\cdots}$ denotes the average obtained via Monte Carlo sampling and $\rho(i\omega_n)$ is the Fourier transformation of $\rho(\tau)=d\sigma_z(\tau)/d\tau$.
From equation~(\ref{eq:path}), $\rho(i\omega_n)$ can be expressed as
\begin{eqnarray}
\rho(i\omega_n) = \sum_{j=1}^{2n}2(-1)^j e^{i\omega_n\tau_j}.
\end{eqnarray}
The susceptibility $\chi(\omega)$ is obtained by the analytical continuation $\chi(\omega)=C(i\omega_n\rightarrow\omega+i\delta)$.
To perform this continuation numerically, we usually employ Pad\'e approximation~\cite{Baker1975,Vidberg1977}.
For the weak coupling regime, Pad\'e approximation yields poor results since the imaginary part of the pole nearest to the real frequency axis is small. 
In this case, we employ another approximation based on the fitting~\cite{Volker1998}.
We assume that the spin correlation function as
\begin{eqnarray}
C(i\omega_n)\approx\frac{a\omega_0^3}{(\omega_n+\lambda)^2+\omega_0^2}+\mathrm{const},
\label{eq:fitting}
\end{eqnarray}
where $a$, $\omega_0$, and $\lambda$ are the fitting parameters determined using the least-squares method.
It is easy to obtain the dynamical susceptibility ${\rm Im}[\chi(\omega)]$ using the fitting function (\ref{eq:fitting}) with optimized parameters.
Note that this fitting method works well for weak couplings since it is compatible with the dynamic susceptibility for the sequential tunneling process.

For using the co-tunneling formula (\ref{eq:cotunneling}), we need to calculate the static susceptibility $\chi_0$.
Typically, a simple estimate $\chi_0 \simeq 2/(\hbar \Delta_{\rm eff})$ yields quantitatively correct results.
However, for the sub-ohmic case, $\chi_0$ has nontrivial temperature dependence, even at low temperatures.
For this case, we numerically calculate $\chi_0$ using the CTQMC method as follows:
\begin{eqnarray}
\chi_0 = \beta\Braket{\bar{\sigma}_z^2}, \\
\bar{\sigma}_z = \frac{1}{\hbar\beta}\int_0^{\hbar\beta}d\tau~\sigma_z(\tau)
= \frac{2}{\hbar\beta}\sum_{j=0}^{2n-1}(-1)^j\tau_j+1.
\end{eqnarray}

\section{Numerical Determination of the Critical Point}
\label{app:Binder}

\begin{figure}[tbp]
	\centering
	\includegraphics[height=11.0cm]{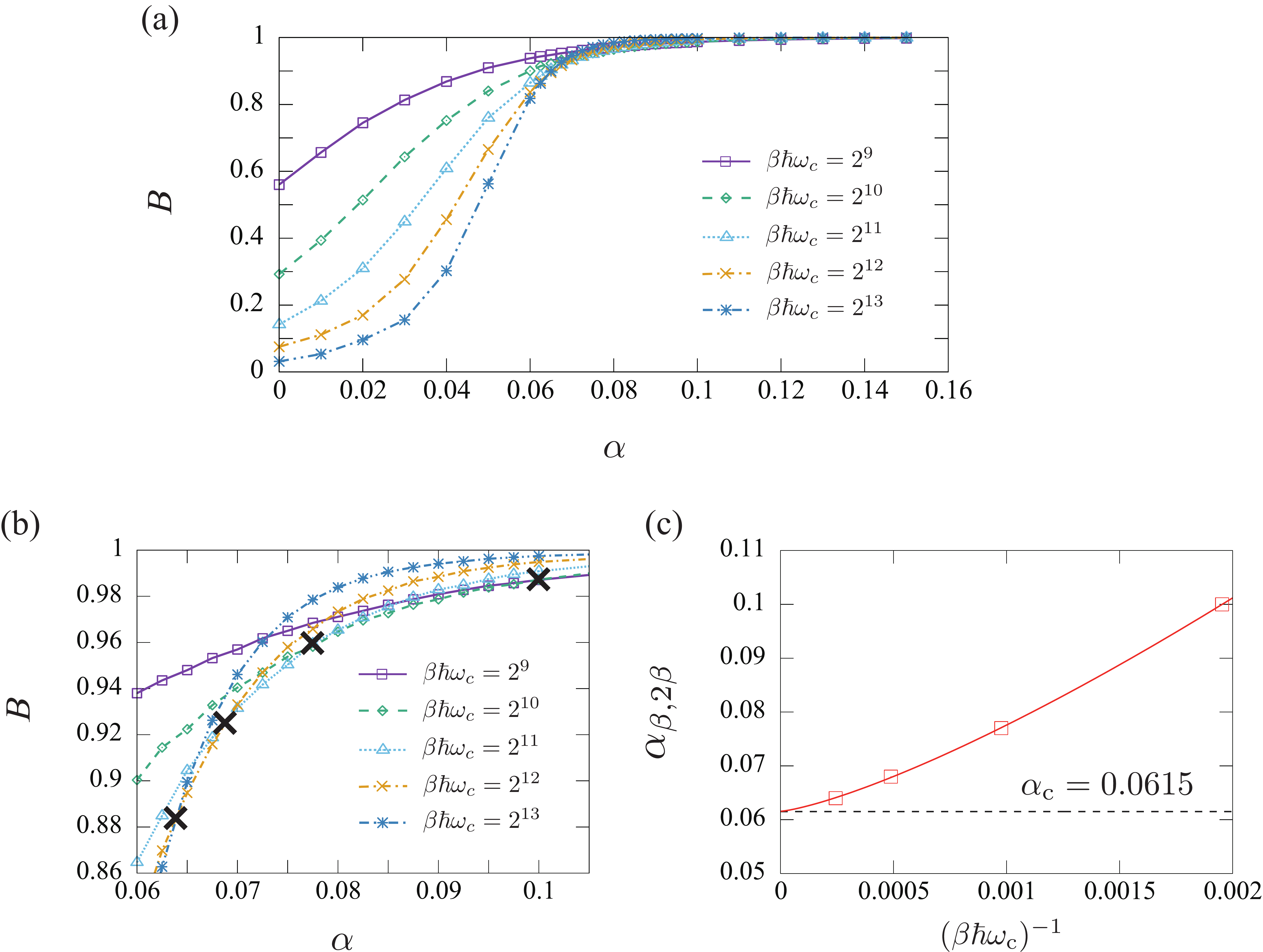}
	\caption{An example of the Binder parameter analysis.
    The results for $s = 0.6$ and $\Delta/\omega_{\rm c} = 0.01$ are shown.
    (a) The Binder parameter as a function of the coupling constant $\alpha$ for different temperatures.
    (b) Enlarged view of (a) around $\alpha = 0.08$. Crosses represent the intersection points $\alpha_{\beta, 2\beta}$ between the two neighboring inverse temperature, $\beta\hbar\omega_c$ and $2\beta\hbar\omega_c$, respectively.
    (c) The intersection points of the Binder parameters. The red solid curve 
    shows the fitted quadratic function.
    The dashed horizontal line indicates the critical value $\alpha_c= 0.0615$
    obtained via the present analysis.}
	\label{fig:Binder}
\end{figure}

In this appendix, we describe how to determine the critical point of the quantum phase transition for the sub-ohmic case ($s<1$).
Following a previous study~\cite{Winter2009}, we introduce the Binder parameter, which is defined as follows:
\begin{eqnarray}
\label{Binder_parameter}
B = \frac{1}{2}\left(3-\frac{\Braket{\bar{\sigma}_z^4}}{\Braket{\bar{\sigma}_z^2}^2}\right), 
\end{eqnarray}
where $\bar{\sigma}_z = (\beta\hbar)^{-1} \int d\tau \sigma_z(\tau)$ and $\langle \cdots \rangle$ indicates the average obtained via the Monte Carlo sampling.
The critical point $\alpha_{\rm c}$ is determined as the point for which the Binder parameter is independent of the temperature at sufficiently low temperatures.
In figure~\ref{fig:Binder}, we show an example of the Binder analysis for $s=0.6$ and $\Delta/\omega_c = 0.01$.
The curve of the Binder parameter for different temperatures has intersection points around $\alpha = 0.08$, as shown in figure~\ref{fig:Binder}~(a).
To accurately determine the critical point, we consider the intersection points $\alpha_{\beta, 2\beta}$ between the two neighboring inverse temperatures, $\beta$ and $2\beta $ [see figure~\ref{fig:Binder}~(b)], and plot the intersection points as a function of $\beta\hbar\omega_c$, as shown in figure~\ref{fig:Binder}~(c).
By extrapolating $\alpha_{\beta, 2\beta}$ in the limit $(\beta\hbar\omega_c)^{-1} \rightarrow 0$ using fitting to the quadratic function of $(\beta\hbar\omega_c)^{-1}$, the critical value $\alpha_{\rm c} = 0.0615$ is obtained for this parameter set.
By performing the same analysis for different values of $s$ and $\Delta/\omega_c$, we finally obtain the phase diagram shown in figure \ref{fig:phase_diagram}.

\section*{References}
\bibliographystyle{iopart-num}
\bibliography{references}

\end{document}